\documentclass[apj]{emulateapj}
\usepackage{color}

\usepackage{ulem}
\usepackage{multirow}

\slugcomment{accepted version (26th April 2017)}
\shortauthors{Sano, Reynoso, Mitsuishi et al.}

\begin{document}
\title{INTERSTELLAR GAS AND X-RAYS TOWARD THE YOUNG SUPERNOVA REMNANT RCW 86;\\PURSUIT OF THE ORIGIN OF THE THERMAL AND NON-THERMAL X-RAY}
\author{H. Sano\altaffilmark{1, 2}, E. M. Reynoso\altaffilmark{3}, I. Mitsuishi\altaffilmark{2}, K. Nakamura\altaffilmark{2}, N. Furukawa\altaffilmark{2}, K. Mruganka\altaffilmark{4}, T. Fukuda\altaffilmark{2}, S. Yoshiike\altaffilmark{2}, A. Nishimura\altaffilmark{2}, A. Ohama\altaffilmark{2}, K. Torii\altaffilmark{5}, T. Kuwahara\altaffilmark{2}, T. Okuda\altaffilmark{6}, H. Yamamoto\altaffilmark{2}, K. Tachihara\altaffilmark{2}, Y. Fukui\altaffilmark{1,2}}
\affil{$^1$Institute for Advanced Research, Nagoya University, Furo-cho, Chikusa-ku, Nagoya 464-8601, Japan; sano@a.phys.nagoya-u.ac.jp}
\affil{$^2$Department of Physics, Nagoya University, Furo-cho, Chikusa-ku, Nagoya 464-8601, Japan}
\affil{$^3$Instituto de Astronom$\mathrm{\acute{\i}}$a y F$\mathrm{\acute{\i}}$sica del Espacio (IAFE, CONICET-UBA), Av. Int. G$\mathrm{\ddot{u}}$raldes 2620, Pabell$\mathrm{\acute{o}}$n IAFE, Ciudad Universitaria, Ciudad Aut$\mathrm{\acute{o}}$noma de Buenos Aires, Argentina}
\affil{$^4$Department of Electrical and Computer Engineering, University of California, San Diego, La Jolla, CA 92093-0407, U.S.A.}
\affil{$^5$Nobeyama Radio Observatory, Minamimaki-mura, Minamisaku-gun, Nagano 384-1305, Japan}
\affil{$^6$National Astronomical Observatory of Japan, Mitaka, Tokyo 181-8588, Japan}

\begin{abstract}
We have analyzed the atomic and molecular gas using the 21 cm H{\sc i} and 2.6/1.3 mm CO emissions toward the young supernova remnant (SNR) RCW 86 in order to identify the interstellar medium with which the shock waves of the SNR interact. We have found an H{\sc i} intensity depression in the velocity range between $-46$ and $-28$ km s$^{-1}$ toward the SNR, suggesting a cavity in the interstellar medium. The H{\sc i} cavity coincides with the thermal and non-thermal emitting X-ray shell. The thermal X-rays are coincident with the edge of the H{\sc i} distribution, which indicates a strong density gradient, while the non-thermal X-rays are found toward the less dense, inner part of the H{\sc i} cavity. The most significant non-thermal X-rays are seen toward the southwestern part of the shell where the H{\sc i} gas traces the dense and cold component. We also identified CO clouds which are likely interacting with the SNR shock waves in the same velocity range as the H{\sc i}, although the CO clouds are distributed only in a limited part of the SNR shell. The most massive cloud is located in the southeastern part of the shell, showing detailed correspondence with the thermal X-rays. These CO clouds show an enhanced CO $J$ = 2--1/1--0 intensity ratio, suggesting heating/compression by the shock front. We interpret that the shock-cloud interaction enhances non-thermal X-rays in the southwest and the thermal X-rays are emitted by the shock-heated gas of density 10--100 cm$^{-3}$. Moreover, we can clearly see an H{\sc i} envelope around the CO cloud, suggesting that the progenitor had a weaker wind than the massive progenitor of the core-collapse SNR RX~J1713.7$-$3949. It seems likely that the progenitor of RCW 86 was a system consisting of a white dwarf and a low-mass star with low-velocity accretion winds.
\end{abstract}


\section{Introduction}
RCW 86 (also known as MSH 14$-$63 or G315.4$-$2.3) is one of the supernova remnants (SNRs) that has been detected in the whole electromagnetic spectrum, from the radio continuum, optical, and infrared domains to the energetic X-rays and GeV/TeV $\gamma$-rays \citep[e.g.,][]{1987A&A...183..118K,1997AJ....114.2664S,2011ApJ...741...96W,2014MNRAS.441.3040B,2016ApJ...819...98A,2016arXiv160104461H}. 
Of particular interest are the bright TeV $\gamma$-rays and non-thermal X-rays, which are tightly related with the production of cosmic-rays (CRs) via the 
diffusive shock acceleration (DSA) mechanism in SNRs \citep{1978MNRAS.182..147B,1978ApJ...221L..29B}. RCW 86 is therefore suitable for studying the origin of Galactic CRs in an energy range $E < 3 \times 10^{15}$ eV and their relationship with the surrounding interstellar medium (ISM) by using multi-wavelength datasets.

RCW 86 is a relatively young SNR, first recorded in AD 185 in the Chinese historical book $Houhanshu$ \citep{1975Obs....95..190C,2006ChJAA...6..635Z}. The SNR is located slightly away from the Galactic plane ($l$, $b$) $\sim$ ($315\fdg4$, $-2\fdg3$), at only 
$\sim2.5$ kpc from us \citep[e.g.,][by association with the edge of the molecular supershell GS 314.8$-$0.1$-$34 discovered by Matsunaga et al. 2001]{1969AJ.....74..879W,1996A&A...315..243R,2013MNRAS.435..910H}. The shell-like morphology of RCW 86 was first discovered 
in radio continuum observations \citep{1961AuJPh..14..497M,1967AuJPh..20..297H}.
After half a century, such morphology has been 
confirmed at all wavelengths, including $\gamma$-rays. The observed diameter is approximately 40 arcmin, corresponding to a diameter of $\sim30$ pc at 2.5 kpc. The progenitor system of RCW 86 (Type Ia or core-collapse (CC)) remains contentious. The CC hypothesis is supported by the presence of several B-type stars in 
the neighbourhood of RCW 86 \citep{1969AJ.....74..879W}. 
However, recent optical and X-ray studies reporting Fe-rich ejecta and
Balmer filaments encircling the shell suggest a Type Ia explosion \citep[e.g.][]{1997AJ....114.2664S,2011PASJ...63S.837Y,2011ApJ...741...96W,2014MNRAS.441.3040B}. 
Besides, RCW 86 lacks a central compact object such as a neutron star or region of O-rich ejecta. Therefore, it is unlikely that RCW 86 is a CC SNR. According to numerical simulations, the progenitor system is also consistent with an off-centered Type Ia explosion \citep[e.g.,][]{2011ApJ...741...96W}.

\begin{figure}
\begin{center}
\includegraphics[width=87mm,clip]{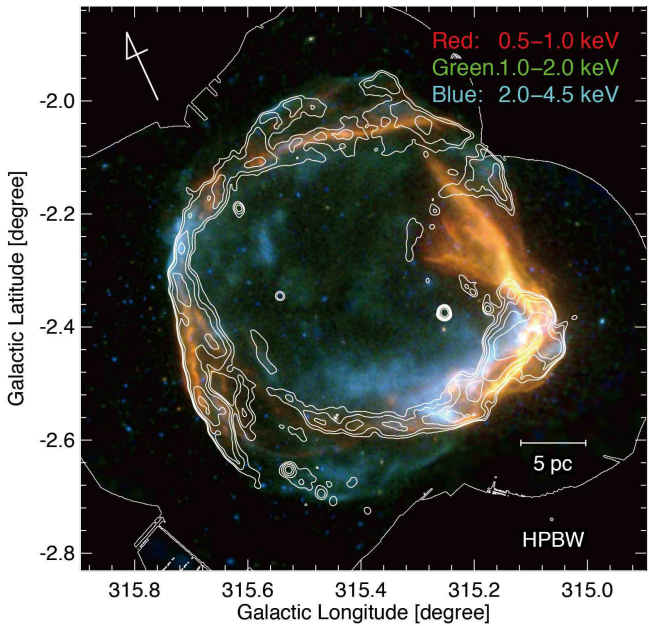}
\caption{Three-color image of the SNR RCW 86 observed by $XMM$-$Newton$. The red, green, and blue colors represent the energy bands, 0.5--1.0, 1.0--2.0, and 2.0--4.5 keV, respectively. The white solid line indicates the region observed with the MOS and PN detectors. Contours represent the MOST radio continuum at a frequency of 843 MHz \citep{1996A&AS..118..329W}. The contour levels are 5, 10, 20, 40, 80, and 160 mJy beam$^{-1}.$}
\label{f01}
\end{center}
\end{figure}

RCW 86 has received much attention since the discovery of TeV $\gamma$-ray emission with the high-energy stereoscopic system (H.E.S.S.) by \cite{2009ApJ...692.1500A}. The TeV $\gamma$-ray flux of RCW 86 is ten times lower than that of the 
Crab nebula, but the origin of which is not yet settled. Subsequently, \cite{2012A&A...545A..28L} and \cite{2014ApJ...785L..22Y} 
obtained GeV $\gamma$-ray images and spectra with the $Fermi$ 
Large Area Telescope (LAT). By using a broad-band spectral energy distribution 
(SED) fitting, they also discussed whether the $\gamma$-rays are hadronic or 
leptonic in origin. They concluded that the leptonic origin was more reasonable,
but the low photon statistics did not rule out a hadronic origin. Recently, 
\cite{2016arXiv160104461H} analyzed the new H.E.S.S. dataset and revealed the 
shell-like morphology in TeV $\gamma$-rays, the origin of which is not yet discerned. Most recently, 
\cite{2016ApJ...819...98A} obtained new GeV $\gamma$-ray images and spectra from
a 6.5-year dataset of $Fermi$ LAT. They concluded that the broad-band SED favors
the leptonic origin under the two-zone model. If the process is hadronic, the 
$\gamma$-rays should spatially correspond to the interstellar gas \citep[e.g.,][]{2008A&A...481..401A,2012ApJ...746...82F,2013ApJ...768..179Y,2014ApJ...788...94F}. Therefore, a detailed spatial comparison between the interstellar gas and $\gamma$-rays is highly desirable in order to establish origin of the high energy emission.

\begin{deluxetable*}{cccccccc}[]
\tablewidth{\linewidth}
\tablecaption{Basic information of $XMM$-$Newton$ observations}
\tablehead{
&&&&&\multicolumn{3}{c}{Exposure}\\
\cline{6-8}\\
Observation ID &  $\alpha_{\mathrm{J2000}}$ & $\delta_{\mathrm{J2000}}$ & Start Date  & End Date & MOS1 & MOS2 & PN \\
& (degree) & (degree) & (yyyy-mm-dd hh:mm:ss) & (yyyy-mm-dd hh:mm:ss) & (ks) & (ks) & (ks)  \\
}
\startdata
0110010701 & 220.73 & $-62.63$ & 2000-08-16 04:04:38 & 2000-08-16 10:43:07 & 17 & 16 & 15\\
0110011301 & 221.31 & $-62.41$ & 2000-08-16 12:03:46 & 2000-08-16 17:37:28 & 11 & 11 & \phantom{0}5\\
0110011401 & 220.51 & $-62.22$ & 2000-08-16 20:18:03 & 2000-08-17 01:36:33 & \phantom{0}9 & 10 & \phantom{0}6\\
0110010501 & 220.14 & $-62.60$ & 2001-08-17 11:47:26 & 2001-08-17 16:25:47 & \phantom{0}9 & \phantom{0}7 & \phantom{0}3\\
0110012501 & 220.24 & $-62.72$ & 2003-03-04 09:46:14 & 2003-03-04 13:11:34 & \phantom{0}8 & \phantom{0}9 & \phantom{0}6\\
0208000101 & 221.26 & $-62.34$ & 2004-01-26 22:30:59 & 2004-01-27 15:12:51 & 46 & 47 & 44\\
0504810101 & 221.57 & $-62.30$ & 2007-07-28 07:45:25 & 2007-07-29 16:12:53 & 95 & 98 & 76\\
0504810601 & 221.57 & $-62.30$ & 2007-07-30 15:45:31 & 2007-07-31 01:52:21 & 18 & 19 & 16\\
0504810201 & 221.40 & $-62.47$ & 2007-08-13 17:42:42 & 2007-08-14 14:37:56 & 50 & 55 & 37\\
0504810401 & 220.15 & $-62.60$ & 2007-08-23 03:17:26 & 2007-08-23 23:33:12 & 62 & 62 & 50\\
0504810301 & 220.50 & $-62.22$ & 2007-08-25 02:49:31 & 2007-08-25 23:34:05 & 61 & 62 & 44\\
0724940101 & 221.22 & $-62.68$ & 2014-01-27 18:48:07 & 2014-01-29 00:03:07 & 96 & 95 & 77
\enddata
\tablecomments{All exposure times correspond to the flare-filtered exposure.}
\label{tab_ex}
\end{deluxetable*}

Studies of the ISM in SNR environments have improved 
our understanding of SNR evolution, shock heating/ionization, acceleration of CRs, and high-energy radiation \citep[e.g.,][]{2012ApJ...746...82F,2012ApJ...744...71I,2013ApJ...768..179Y,2013ApJ...778...59S}. In RCW 86, however, deep studies of the ISM have not
been reported. \cite{2011ApJ...741...96W} revealed the interstellar dust distribution of RCW 86 using the $Spitzer$ $Space$ $Telescope$ and the $Wide$-$Field$ 
$Infrared$ $Survey$ $Explorer$ ($WISE$). They noted the distribution of thin 
dust filaments in the east region, which appear to trace the SNR shockwaves. In contrast, neutral atomic gas (H{\sc i}) forms a cavity-like structure at radial velocities of approximately $-34$ km s$^{-1}$ \citep{2016ApJ...819...98A,2016BAAA...58..212D}, although the detailed velocity structure and its 
relationship with the SNR shockwaves have not been presented. In
particular, observations of molecular clouds traced by carbon monoxide (CO) 
emission have not been attempted to date. In proper-motion measurements, the shock velocity was found to differ from region to region perhaps owing to the inhomogeneous interstellar environment and/or different stages of interaction with the surroundings \citep[e.g.,][]{1997A&A...328..628V,2013MNRAS.435..910H}. The highest shock velocity ($\sim3,000$ km s$^{-1}$) occurs in the northeast region, which mainly comprises non-thermal X-rays \citep{2013MNRAS.435..910H,2016ApJ...820L...3Y}. Conversely, the lowest shock velocities (500--900 km s$^{-1}$) are observed in the southwest and northwest regions, which strongly emit thermal X-rays \citep{1990ApJ...358L..13L,2001ApJ...547..995G}. Moreover, according to \cite{2002ApJ...581.1116R} and \cite{2011PASJ...63S.837Y}, interactions between the dense clouds and SNR shockwaves manifest as reverse or secondary shocks in some parts of the shell.

In the present study, we aim to identify the interstellar molecular/atomic gas distribution associated with RCW 86 and to compare it with the thermal/non-thermal X-rays, radio continuum, and H$\alpha$ datasets. We seek for the physical connection between the surrounding gas components, and pursuit the origin of the thermal/non-thermal X-rays, shock properties, and the progenitor system of the SNR. In a subsequent paper, we will compare the shock-interacting gas and TeV $\gamma$-rays (Sano et al. 2017, in preparation). Section 2 presents the observations and data reduction of NANTEN2 CO, ATCA $\&$ Parkes H{\sc i}, $XMM$-$Newton$ X-rays, and the datasets at the other wavelengths. Section 3 comprises four subsections. Subsection 3.1 overviews the CO, H{\sc i}, and X-ray distributions; subsections 3.2 and 3.3 present a 
detailed analysis of the distributions and physical conditions of CO and H{\sc i} respectively; and subsection 3.4 presents a detailed comparison between these and the X-ray distributions. Discussion and conclusions are presented in Sections 4 and 5, respectively.

\section{OBSERVATIONS $\&$ DATA REDUCTIONS}
\subsection{CO}
We performed $^{12}$CO($J$ = 1--0, 2--1) observations with NANTEN2 4 m millimeter/sub-millimeter telescope at Pampa la Bola in northern Chile (4,865 m above sea level).
Observations of the $^{12}$CO($J$ = 1--0) emission line at 115 GHz were conducted from December 2012 to January 2013. The front end was a 4-K cooled superconductor-insulator-superconductor (SIS) mixer receiver. The double-sideband (DSB) system temperature was $\sim110$ K toward the zenith including the atmosphere. The back end was a digital Fourier transform spectrometer (DFS) with 16,384 channels of 1 GHz bandwidth, corresponding to a velocity coverage of $\sim2,600$ km s$^{-1}$. Frequency and velocity resolutions were 61 kHz and $\sim0.16$ km s$^{-1}$ ch$^{-1}$, respectively. We used the on-the-fly (OTF) mode with Nyquist sampling, and the observed area was one square degree. After convolving the datacube with a Gaussian kernel of $\sim90$ arcsec (FWHM), the typical noise level was 0.42 K ch$^{-1}$. The final beam size was $\sim180$ arcsec (FWHM). The pointing accuracy was checked every 3 hours. An offset better than $\sim$25 arcsec was achieved. The absolute intensity was calibrated by observing IRAS 16293$-$2422 [$\alpha$(J2000) = $16^{\mathrm{h}}32^{\mathrm{m}}23.3^{\mathrm{s}}$, $\delta$(J2000) = $-24{^\circ}28\arcmin39\farcs2$] \citep{2006AJ....131.2921R}.

\begin{figure*}
\begin{center}
\includegraphics[width=168mm,clip]{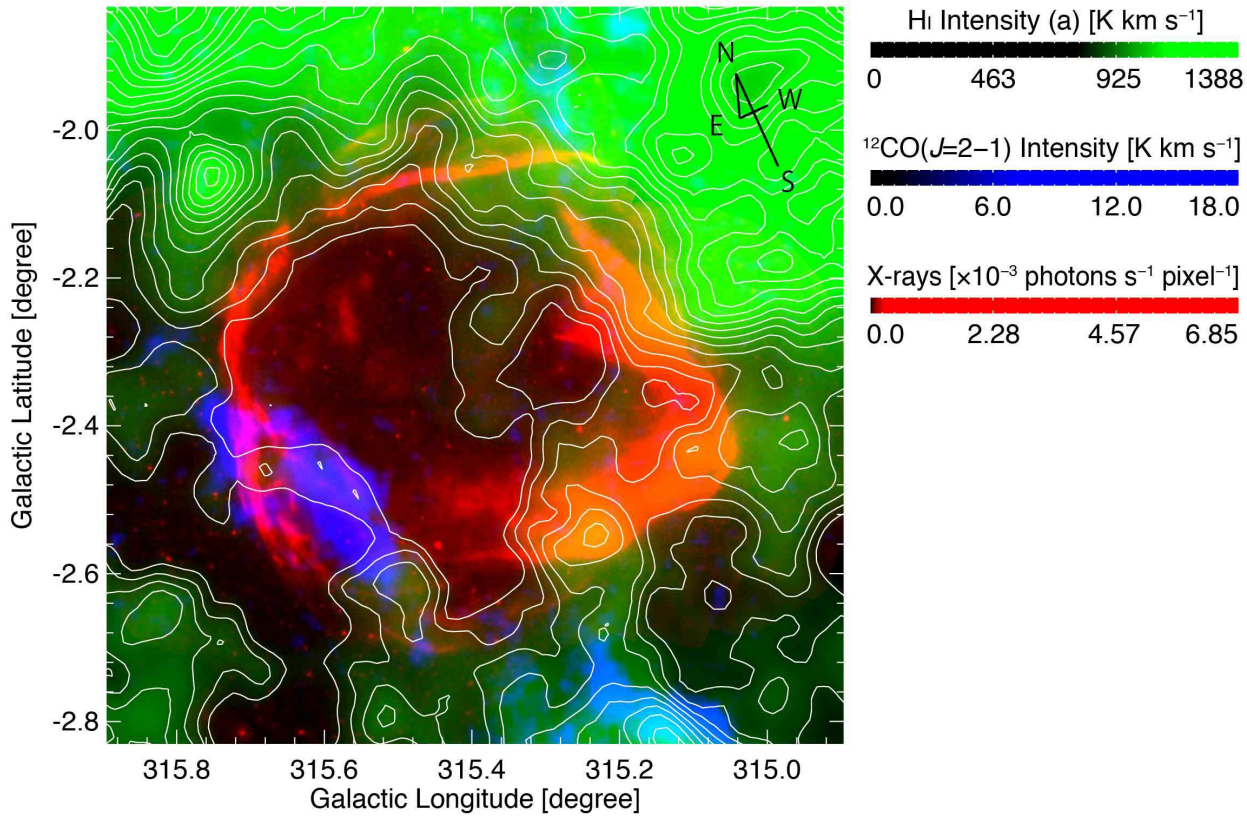}
\caption{Three-color images of the SNR RCW 86 and its surroundings. The red, blue, and green colors represent the $XMM$-$Newton$ broad-band X-rays (0.5--4.5 keV), NANTEN2 $^{12}$CO($J$ = 2--1), and the ATCA $\&$ Parkes H{\sc i}, respectively. The velocity range of CO and H{\sc i} is from $-46.0$ to $-28.0$ km s$^{-1}$. The contours indicate the H{\sc i} integrated intensity. The lowest contour level and the contour interval are 806.4 K km s$^{-1}$ ($\sim 240 \sigma$) and 33.6 K km s$^{-1}$ ($\sim 10 \sigma$), respectively. }
\label{f02}
\end{center}
\end{figure*}%

\begin{figure*}
\begin{center}
\includegraphics[width=168mm,clip]{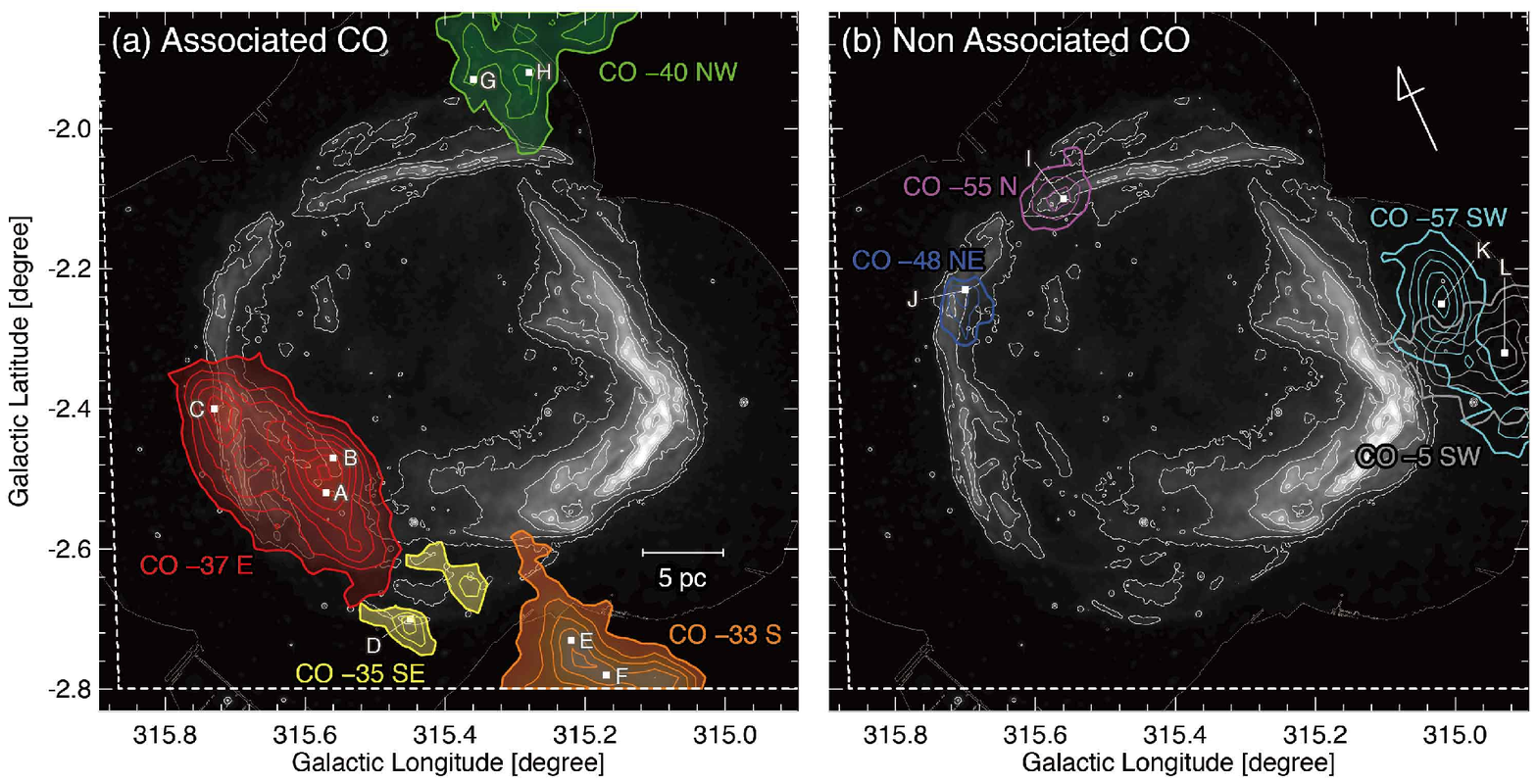}
\caption{Maps of the (a) associated and (b) non-associated $^{12}$CO($J$ = 1--0) clouds shown by colored contours. White contours indicate the X-ray intensity in the energy band from 0.5 to 4.5 keV. The contour levels are 6.00$\times$10$^{-5}$, 1.42$\times$10$^{-4}$, 3.87$\times$10$^{-4}$, 7.95$\times$10$^{-4}$, 1.37$\times$10$^{-3}$, 2.10$\times$10$^{-3}$, and 3.00$\times$10$^{-3}$ photons s$^{-1}$ pixel$^{-1}$. The integration velocity ranges are as follows: $-39.0$ to $-34.4$ km s$^{-1}$ for CO $-37$ E (contours: lowest $\sim$1.4 K km s$^{-1}$, intervals $\sim$2.6 K km s$^{-1}$), $-34.0$ to $-35.6$ km s$^{-1}$ for CO $-35$ SE (contours: lowest $\sim$0.9 K km s$^{-1}$, intervals $\sim$0.6 K km s$^{-1}$), $-35.9$ to $-29.3$ km s$^{-1}$ for CO $-33$ S (contours: lowest $\sim$1.8 K km s$^{-1}$, intervals $\sim$2.5 K km s$^{-1}$), $-44.3$ to $-35.2$ km s$^{-1}$ for CO $-40$ NW (contours: lowest and intervals $\sim$2.1 K km s$^{-1}$), $-56.0$ to $-54.4$ km s$^{-1}$ for CO $-55$ N (contours: lowest $\sim$0.9 K km s$^{-1}$, intervals $\sim$1.6 K km s$^{-1}$), $-48.6$ to $-46.5$ km s$^{-1}$ for CO $-48$ NE (contours: lowest $\sim$0.7 K km s$^{-1}$, intervals $\sim$1.1 K km s$^{-1}$), $-57.8$ to $-55.7$ km s$^{-1}$ for CO $-57$ SW (contours: lowest $\sim$1.1 K km s$^{-1}$, intervals $\sim$1.8 K km s$^{-1}$), and $-4.9$ to $-4.1$ km s$^{-1}$ for CO $-5$ SW (contours: lowest and intervals $\sim$0.8 K km s$^{-1}$). The positions of CO peaks detected at different radial velocities (see Table \ref{tab1}) are identified by letters (A--L).}
\label{f03}
\end{center}
\end{figure*}%

Observations of the $^{12}$CO($J$ = 2--1) emission line at 230 GHz were conducted in November 2008. The front end was a 4-K cooled SIS mixer. The system temperature in DSB was $\sim120$ K toward the zenith including the atmosphere. We used an acousto-optical spectrometer with 2,048 channels of 250 MHz bandwidth corresponding to a velocity coverage of $\sim385$ km s$^{-1}$. The frequency and velocity resolutions were 250 kHz and $\sim0.38$ km s$^{-1}$ ch$^{-1}$, respectively. We used the OTF mode with Nyquist sampling, and the observed area was $\sim0.88$ square degrees. After convolving the datacube with a Gaussian kernel of $\sim45$ arcsec (FWHM), the typical one sigma noise fluctuations were less than 0.3 K ch$^{-1}$. The final smoothed beam size was $\sim90$ arcsec (FWHM). The pointing error was less than 15 arcsec, and the intensity calibration was applied by observing Oph EW4 [$\alpha$(J2000) = $16^{\mathrm{h}}26^{\mathrm{m}}21.92^{\mathrm{s}}$, $\delta$(J2000) = $-24{^\circ}25\arcmin40\farcs4$(J2000)] \citep{2005ApJ...625..194K}.

\subsection{\rm H{\sc i}}
We performed H{\sc i} observations at 1420 MHz using the Australia Telescope Compact Array (ATCA), which consists of six 22-m dishes located at Narrabri, Australia. Observations were conducted during 13 hours on March 24--25, 2002, with the ATCA in the EW 367 configuration (baselines from 46 to 367 m, or from 0.3 to 1.75 k$\lambda$, excluding the 6$^{\rm th}$ antena). We employed the mosaicking technique, with 45 pointings covering an area of $\sim 4$ square degrees. The absolute flux density scale was determined by observing PKS B1934$-$638, which was used as the primary amplitude and bandpass calibrator. We also periodically observed PKS 1352$-$63 for gain and phase calibration. Data reduction was performed by using the MIRIAD software package \citep{1995ASPC...77..433S}. The images were retrieved using a superuniform weighting and keeping only visibilities shorter than 1.1 k$\lambda$. To include extended emission, we combined the ATCA data-set with single-dish observations performed with the Parkes 64 m telescope. The final beam size is 160 arcsec $\times$ 152 arcsec with a position angle of $-3^\circ$. Typical noise level is 1.0 K at 0.82 km s$^{-1}$ velocity resolution. The data are identical to those published by \citet{2016ApJ...819...98A}.

\subsection{X-rays}

Twelve $XMM$-$Newton$ pointed observation data are available for RCW~86 as 
summarized in Table \ref{tab_ex}. We analyzed both the EPIC-pn and EPIC-MOS 
datasets by using the HEAsoft version 6.18 and the $XMM$-$Newton$ Science 
Analysis System (SAS) version 15.0. We reprocessed the observation data files 
following standard procedures provided by the $XMM$-$Newton$ extended source 
analysis software \citep[ESAS,][]{2008A&A...478..575K}. In order to create 
instrumental background-subtracted, exposure-corrected, adaptively-smoothed 
images, we prepared exposure maps and quiescent particle background (QPB) images
for each observation by using the mos-/pn-filter and mos-/pn-back tasks. Then, 
we combined the images after subtracting the QPB, and the combined images were 
divided by the merged exposure maps. An adaptive smoothing process was also 
applied to emphasize diffuse components with the pixel size of $4''$. Finally, 
we obtained QPB-subtracted, exposure-corrected, adaptively-smoothed images in 
the energy bands of 0.5--1.0 (soft) /1.0--2.0 (middle) / 2.0--4.5 (hard) / 
0.5--4.5 (broad) keV. In this analysis, high background periods are removed and 
the net exposure time is shown in Table \ref{tab_ex}. 
Figure \ref{f01} shows an X-ray tricolor image of RCW 86. The red, green, and blue regions emit at 0.5--1.0 keV (soft band), 1.0--2.0 keV (medium band), and 2.0--4.5 keV (hard band), respectively. The soft and hard bands are dominated by continuum radiation from thermal plasma and synchrotron X-rays produced by the TeV CR electrons, respectively \citep{2002ApJ...581.1116R,2016ApJ...819...98A}. 

In the present paper, we shall hereafter refer to the emission seen in the image in the soft band as ``thermal X-rays'' and that of the hard band as ``non-thermal X-rays'' because each energy band is dominated by the continuum radiation from thermal plasma and non-thermal synchrotron X-rays, respectively \citep[e.g.,][]{2002ApJ...581.1116R,2016ApJ...819...98A}. Moreover, thermal X-rays are dominated by the ISM plasma components, whose distribution is significantly different from the ejecta component except for the SW region \citep{2011PASJ...63S.837Y}.

\subsection{Astronomical Data at the Other Wavelengths}
H$\alpha$ and radio continuum data are used to derive the spatial distribution of the ionized gas and low-energy CR electrons. We used the H$\alpha$ and 843 MHz radio continuum data that appear in the Southern H-Alpha Sky Survey Atlas \citep[SHASSA;][]{2001PASP..113.1326G} and the Molonglo Observatory Synthesis Telescope (MOST) Supernova Remnant Catalogue \citep[MSC;][]{1996A&AS..118..329W}, in addition to the CO/H{\sc i} and X-ray data. The angular resolutions of H$\alpha$ and radio continuum are 48 arcsec and 43 arcsec, respectively.

\section{RESULTS}
\subsection{Overview of CO, H{\sc i}, and X-ray Distributions}

To determine the velocity range of the atomic and molecular gas associated with the SNR RCW 86, we carried out the following steps:

\begin{enumerate}
\item Searching by visual inspection for a good spatial correspondence between the ISM and X-ray intensities in the velocity channel distribution of CO/H{\sc i} overlaid upon the X-ray contours (see Appendix and Figure \ref{fa1});
\item Investigating the physical conditions of associated molecular clouds using the $^{12}$CO $J$ = 2--1/1--0 intensity ratio maps (see Section 3.2);
\item Exploring possible evidences of expanding motions of H{\sc i} and CO due to the SNR shockwaves and/or stellar winds from the progenitor of RCW 86 (see Section 3.3).
\end{enumerate}

This analysis led us to conclude that the gas associated with RCW 86 is most likely found at a velocity range from $-46$ to $-28$ km s$^{-1}$. The ATCA $\&$ Parkes H{\sc i} and NANTEN2 $^{12}$CO($J$ = 2--1) emissions integrated in this velocity range are displayed in green and blue, respectively, in Figure \ref{f02}, together with the $XMM$-$Newton$ X-ray image (red: 0.5--4.5 keV) of RCW 86. 

Towards the north, there is an H{\sc i} intensity gradient increasing from east to west, and the most prominent features, with intensities above 1,000 K km s$^{-1}$, lie in the northwest region. The overall distribution of the H{\sc i} clouds 
tend to encircle the X-ray shell-like structure. We also find that the diffuse 
H{\sc i} gas, with an intensity of $\sim700$ K km s$^{-1}$, fills the interior 
of the SNR shell. To the east, a large CO cloud with diffuse H{\sc i} emission 
is located toward the X-ray filaments. The high angular resolution of CO 
allowed us to see that the X-ray emission of the filament located around ($l$, 
$b$) $\sim$ ($315\fdg7$, $-2\fdg5$) is higher where the emission of the CO cloud is lower.
The CO clouds are located not only at the east but also at the south and the northwest. Four additional CO clouds are visible toward the SNR: CO $-$57 SW, CO $-$55 N, CO $-$48 NE, and CO $-$5 SW (Figure \ref{f03}b). These are probably not interacting with the SNR because their radial velocities do not coincide with those of the associated CO clouds. Hereafter, we shall focus on the velocity range from $-$46 to $-$28 km s$^{-1}$ which contains the associated CO clouds.

\begin{deluxetable*}{lccccccccccc}
\tabletypesize{\scriptsize}
\tablecaption{Physical Properties of $^{12}$CO($J$ = 1--0) Clouds}
\tablewidth{0pt}
\tablehead{
\multicolumn{1}{c}{Name}  & Position & $l$ & $b$ & $T_R^\ast $ & $V_{\mathrm{peak}}$ &  $\Delta V_{\mathrm{LSR}}$ & $N_\mathrm{p}$(H$_2$) & Size  & Mass & Associated\\
 & &(deg) & (deg) & (K) & (km $\mathrm{s^{-1}}$) & (km $\mathrm{s^{-1}}$) & ($\times 10^{21}$ cm$^{-2}$) & (pc) & ($M_\sun $) & \\
\multicolumn{1}{c}{(1)} & (2) & (3) & (4) & (5) & (6) & (7) & (8) & (9) & (10) & (11)
}
\startdata
\multirow{3}{*}{CO $-37$ E} & A & 315.57 & $-2.52$ & 6.9 & $-35.5$ & 2.2 & 5.7 & \multirow{3}{*}{12.4} & \multirow{3}{*}{3699} & \multirow{3}{*}{Yes} \\
& B & 315.56 & $-2.47$ & 4.2/3.5 & $-35.7$/$-37.6$ & 2.0/2.8 & 5.8\\
& C & 315.73 & $-2.40$ & 5.6 & $-36.3$ & 2.4 & 5.3\\
\hline
CO $-35$ SE & D & 315.43 & $-2.70$ & 2.0 & $-34.7$ & 1.6 & 3.4 & 5.1 & 170 & Yes \\
\hline
\multirow{2}{*}{CO $-33$ S} & E & 315.22 & $-2.73$ & 2.7 & $-33.7$ & 4.6 & 4.6 & \multirow{2}{*}{$>$8.6} & \multirow{2}{*}{$>$1520} & \multirow{2}{*}{Yes} \\
 & F & 315.17 & $-2.78$ & 3.8 & $-30.8$ & 3.0 & 4.5 & \\
\hline
\multirow{2}{*}{CO $-40$ NW} & G & 315.36 & $-1.93$ & 1.9 & $-35.8$ & 1.3 & 1.9 & \multirow{2}{*}{$>$9.0} & \multirow{2}{*}{$>$1070} & \multirow{2}{*}{Yes} \\
 & H & 315.28 & $-1.92$ & 2.6 & $-43.0$ & 2.4 & 2.9 & \\
\hline
CO $-55$ N & I & 315.36 & $-2.10$ & 3.3 & $-55.1$ & 1.7 & 1.9 &  &  & No \\
\hline
CO $-48$ NE & J & 315.70 & $-2.23$ & 1.6 & $-47.5$ & 1.9 & 1.1 &  &  & No \\
\hline
CO $-57$ SW & K & 315.02 & $-2.25$ & 5.2 & $-56.7$ & 2.0 & 3.8 &  &  & No \\
\hline
CO\phantom{0} $-5$ SW & L & 314.93 & $-2.32$ & 4.5 & \phantom{0}$-4.5$ & 0.8 & 1.4 &  &  & No
\enddata
\label{tab1}
\tablecomments{Col. (1): Cloud name. Col. (2): Position name. Cols. (3--4): Position of the maximum CO intensity for each velocity component. Cols. (5--8): Physical properties of the $^{12}$CO($J$ = 1--0) emission obtained at each position. Col. (5): Peak radiation temperature, $T_R^{\ast} $. Col. (6): $V_{\mathrm{peak}}$ derived from a single Gaussian fitting. Col. (7): Full-width half-maximum (FWHM) line width, $\bigtriangleup V_{\mathrm{peak}}$. Col. (8): Proton column density $N_\mathrm{p}$(H$_2$) derived from the CO integrated intensity, $W$($^{12}$CO), $N$($\mathrm{H_2}$) = 2 $\times $  $10^{20}$ [$W$($^{12}$CO)/(K km $\mathrm{s^{-1}}$)] ($\mathrm{cm^{-2}}$) \citep{1993ApJ...416..587B}. Col. (9): Cloud size defined as ($A$/$\pi$)$^{0.5} \times 2$, where $A$ is the total cloud surface area surrounded by the 3 $\sigma$ levels in the integrated intensities of each CO cloud. Col. (10): Mass of the cloud derived using the relation between the molecular hydrogen column density $N$($\mathrm{H_2}$), and the $^{12}$CO($J$ = 1--0) integrated intensity, $W$($^{12}$CO), shown in Col. (8).}
\end{deluxetable*}

\begin{figure*}[h]
\begin{center}
\includegraphics[width=168mm,clip]{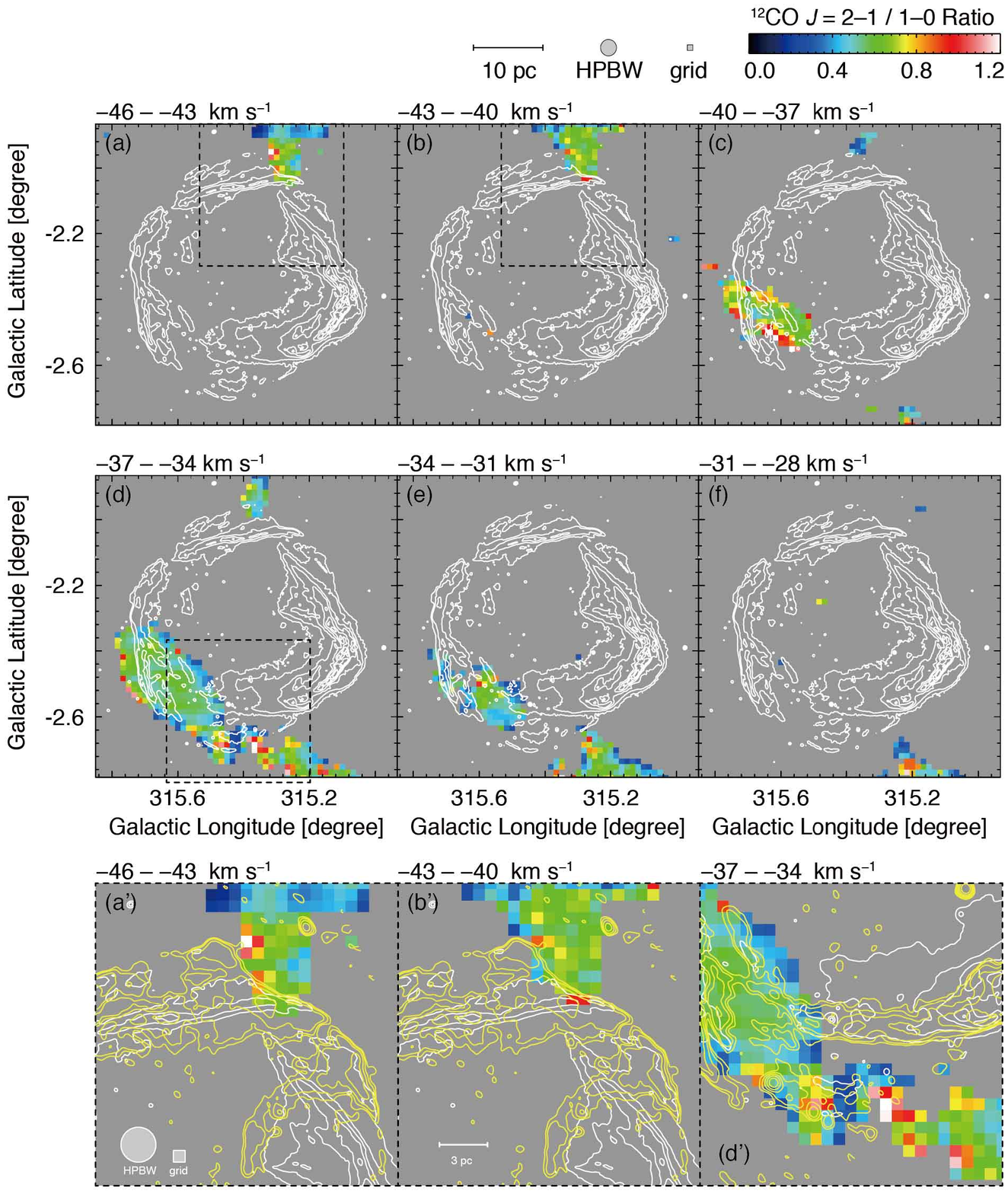}
\caption{(a--f) Velocity channel maps of the line intensity ratio $^{12}$CO $J$ = 2--1/1--0 at an interval of 3 km s$^{-1}$ in a velocity range from $-46.0$ to $-28.0$ km s$^{-1}$. The white contours show the X-ray intensity distributions shown in Figure \ref{f03}. (a$'$, b$'$, d$'$) Enlarged views toward dashed regions in Figures \ref{f04}a, b, and d. The yellow contours indicate the MOST radio continuum at a frequency of 843 MHz. The contour levels are 2.5, 5, 10, 20, 40, 80, and 160 mJy beam$^{-1}.$}
\label{f04}
\end{center}
\end{figure*}%

Figure \ref{f03}a shows the distribution of the molecular clouds associated 
with the SNR. These clouds are named CO $-40$ NW, CO $-37$ E, CO $-35$ SE, and 
CO $-33$ S, respectively, and their peak radial velocities are derived from a single Gaussian fitting. The basic physical properties of the CO clouds are listed in Table \ref{tab1}. All physical parameters were estimated based on a distance of 2.5 kpc. The CO clouds are at the same distance since they have similar radial velocities around $\sim -35$ km s$^{-1}$. We see that there are no broad-line features with velocity-widths above 10 km s$^{-1}$ in the CO spectra. In order to estimate the mass of the CO clouds $M_\mathrm{cloud}$, we used the following equation: 

\begin{eqnarray}
M_{\mathrm{cloud}} = \mu m_{\mathrm{H}} \sum_{i} [D^2 \Omega N_i(\mathrm{H}_2)],
\label{eq1}
\end{eqnarray}

where $\mu$ is the mean molecular weight, $m_{\mathrm{H}}$ is the mass of the atomic hydrogen, $D$ is the distance to the CO cloud, $\Omega$ is the solid angle of a square pixel, and $N_i(\mathrm{H}_2)$ is the hydrogen column density of each pixel $i$ in the Galactic longitude-latitude plane. We used $\mu$ = 2.8 to account for a helium abundance of 20 $\%$. The hydrogen column density $N(\mathrm{H}_2)$ is derived by using the relationship 

\begin{eqnarray}
X = N(\mathrm{H}_2) / W(^{12}\mathrm{CO}),
\label{eq2}
\end{eqnarray}

where $X$ is an X-factor in units of cm$^{-2}$ (K km s$^{-1}$)$^{-1}$. We used $X$ = 2.0 $\times$ 10$^{20}$ in the present paper \citep{1993ApJ...416..587B}. We estimated the total mass of the CO clouds to be at least $\sim$6,500 $M_{\odot}$. 

\begin{figure*}
\begin{center}
\includegraphics[width=168mm,clip]{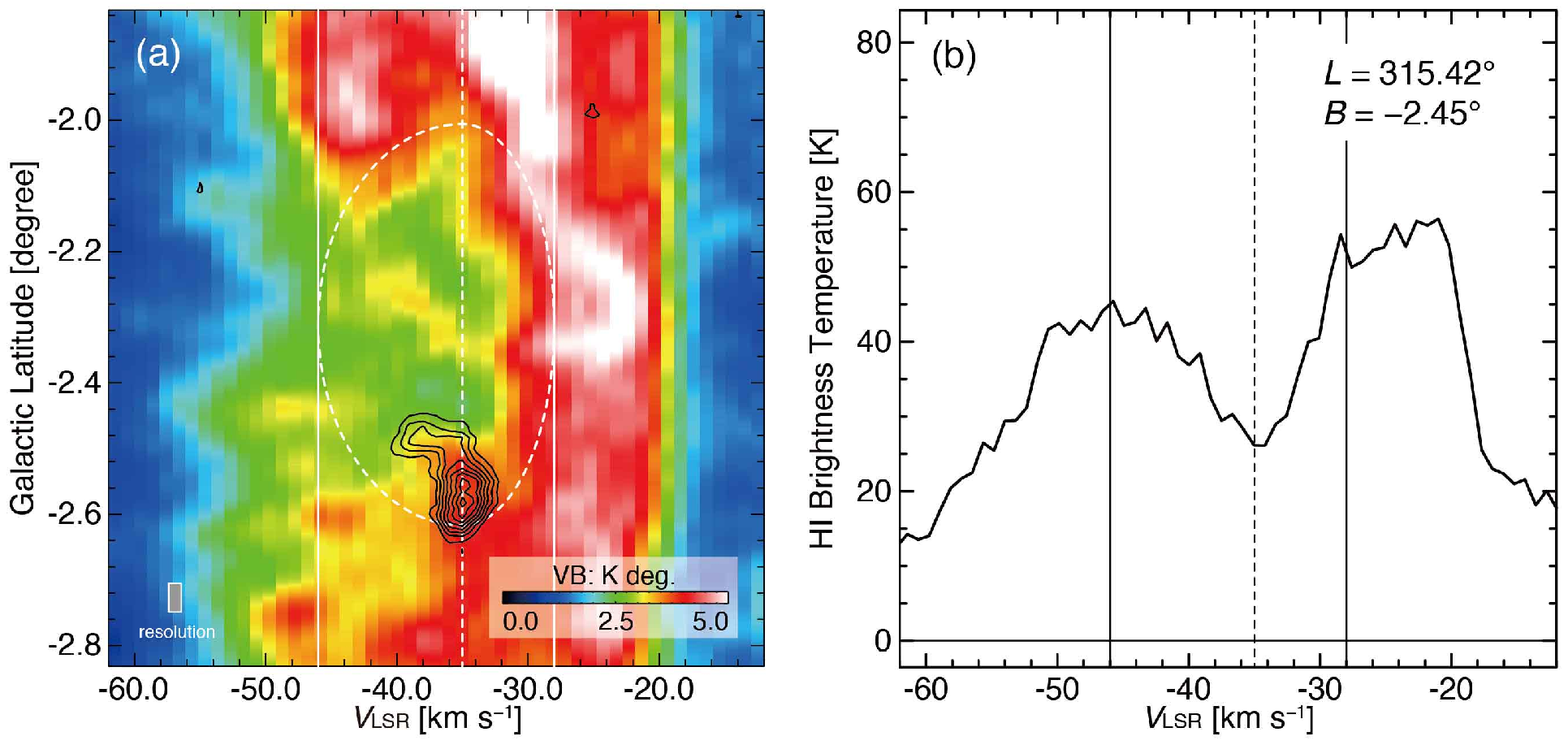}
\caption{(a) Velocity-Latitude diagram of H{\sc i}. The integration range in Galactic longitude is from $315\fdg48$ to $315\fdg56$, as shown in Figure \ref{f02}. Black contours indicate the intensity distribution of $^{12}$CO($J$ = 2--1). The lowest contour level and intervals are 0.03 K degree and 0.02 K degree, respectively. Dashed circle and vertical solid lines show an asymmetrically expanding spherical shell and the velocity integration range of Figure \ref{f02}. (b) H{\sc i} spectra at ($l$, $b$) = ($315\fdg42$, $-2\fdg45$). Velocity range and vertical lines are the same as Figure \ref{f05}a.}
\label{f05}
\end{center}
\end{figure*}%

\subsection{Physical Conditions of Molecular Gas}

In order to investigate the physical conditions of the associated CO clouds, we 
have calculated the line intensity ratio map using the $^{12}$CO($J$ = 
2--1) and $^{12}$CO($J$ = 1--0) emission lines. The intensity ratio corresponds 
to the degree of the rotational excitation of molecules, which reflects the gas 
density and/or temperature. Both datasets were smoothed to 
an angular resolution of $\sim$180 arcsec (FWHM) and summed up to 1 km s$^{-1}$ 
per velocity bin. The data points used for the analysis were those above 
the $3\sigma$ noise level in both lines.

Figure \ref{f04} shows the velocity channel distributions of the line intensity 
ratio $^{12}$CO $J$ = 2--1/1--0 every 3 km s$^{-1}$. We found that part of the 
CO $-37$ E cloud shows an intensity ratio significantly higher  than 0.8 (Figure \ref{f04}c), while the region in 
the immediate vicinity of the cloud shows values smaller than 0.6 (Figures \ref{f04}c, \ref{f04}d, and \ref{f04}e). This may be 
due to some external influences that affect only the surface of 
the clouds because an intensity ratio of $<$ 0.6 is typical 
of dark molecular clouds in the Milky Way 
without extra heating \citep[e.g.,][]{1997ApJ...486..276S}. In addition to the 
CO $-37$ E cloud, we note that the edges of the CO $-40$ NW, CO $-35$ SE, and CO
$-33$ S clouds also have intensity ratios 
higher than 0.8. Figures \ref{f04}a$'$, b$'$, and d$'$ show the line intensity ratio maps toward these clouds superposed with 
the same radio continuum contours as in Figure \ref{f01}. The regions having intensity ratios higher than 0.8 are located along the radio shell of RCW 86. This is not considered to be 
due to stellar feedback since there are 
no $IRAS$/$AKARI$ infrared point sources or OB type stars in these regions \citep[e.g.,][]{1969AJ.....74..879W,1988iras....7.....H,2010A&A...514A...1I}. Therefore, this enhanced ratio indicates shock heating/compression due to the forward shock and/or stellar winds from the progenitor of RCW 86, which supports the association between the SNR and the CO clouds. 

\begin{figure*}
\begin{center}
\includegraphics[width=166mm,clip]{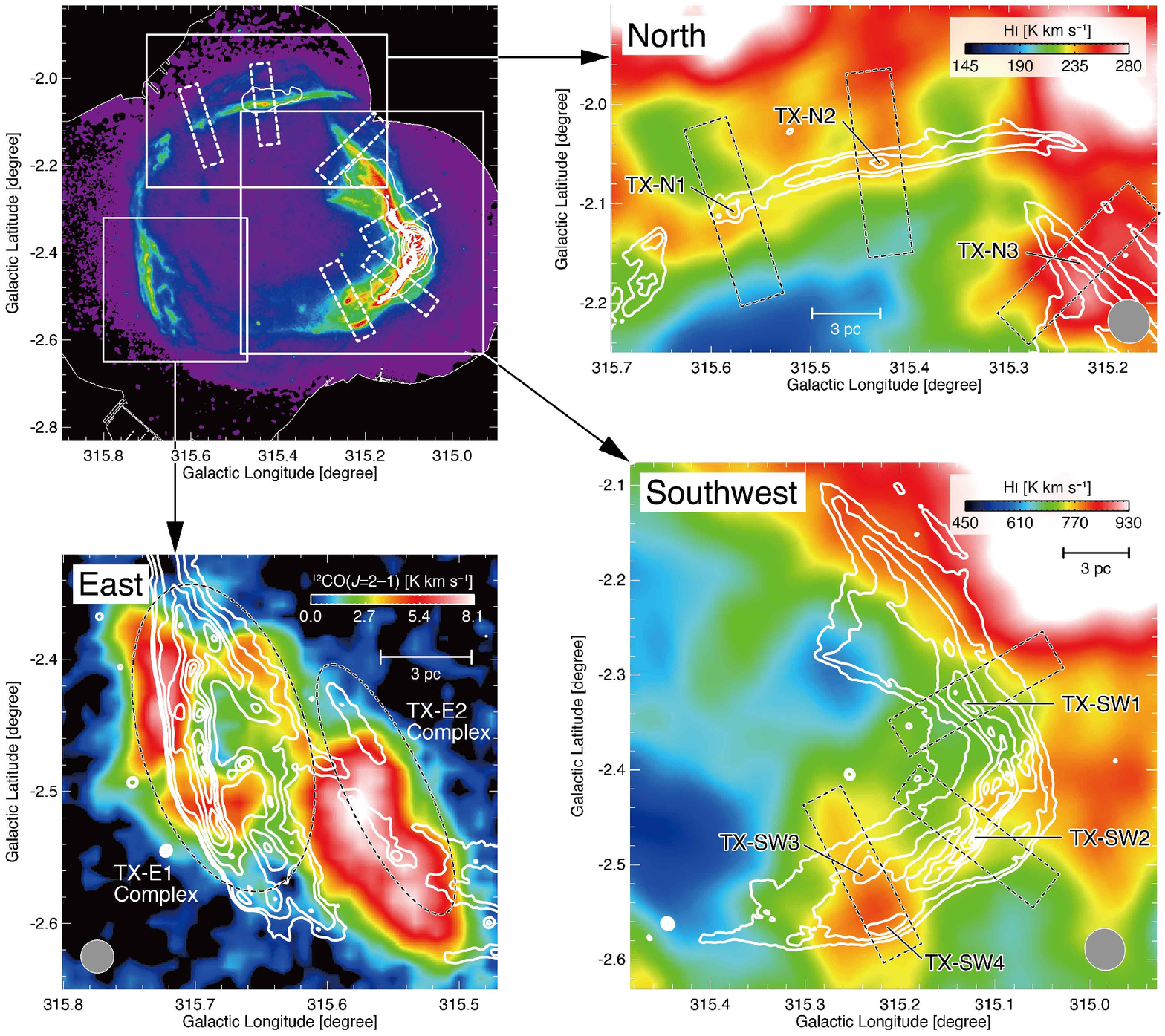}
\caption{Top left: $XMM$-$Newton$ X-ray image in the energy band of 0.5--1.0 keV overlaid with the H$\alpha$ intensity contours (lowest: 500 dR, interval: 250 dR) taken from SHASSA \citep{2001PASP..113.1326G}. Other panels: Distributions of H{\sc i} and $^{12}$CO($J$ = 2--1) obtained with ATCA $\&$ Parkes and NANTEN2 (rainbow scale) superposed with the $XMM$-$Newton$ X-ray contours in the energy band of 0.5--1.0 keV. The three regions toward the north, east, and southwest are shown with inserts in the X-ray image on the top left panel. The velocity range spans from $-32$ to $-29$ km s$^{-1}$ in the north; from $-36$ to $-34$ km s$^{-1}$ in the east; and from $-42$ to $-28$ km s$^{-1}$ in the southwest. The contour levels of the X-rays are 6.00$\times$10$^{-5}$, 1.42$\times$10$^{-4}$, 3.87$\times$10$^{-4}$, 7.95$\times$10$^{-4}$, 1.37$\times$10$^{-3}$, 2.10$\times$10$^{-3}$, and 3.00$\times$10$^{-3}$ photons s$^{-1}$ pixel$^{-1}$ for the north and the southwest; 2.50$\times$10$^{-5}$, 3.54$\times$10$^{-5}$, 6.67$\times$10$^{-5}$, 1.19$\times$10$^{-4}$, 1.92$\times$10$^{-4}$, 2.85$\times$10$^{-4}$, and 4.00$\times$10$^{-4}$ photons s$^{-1}$ pixel$^{-1}$ for the east.}
\label{f06}
\end{center}
\end{figure*}%

\subsection{Expanding Structure and Physical Properties of {\rm H}{\sc i} and CO}
Figure \ref{f05}a shows the H{\sc i} velocity-latitude diagram. The integration 
range in Galactic longitude is from $315\fdg48$ to $315\fdg56$, as shown in 
Figure \ref{f02}. We found an H{\sc i} cavity-like structure in the radial velocity range from $-46$ to $-28$ km s$^{-1}$, which has a size similar to RCW 86 in terms of the Galactic latitude range ($\sim40$ arcmin; $\sim30$ pc at the distance of 2.5 kpc). 
The large velocity range involved, nearly 20 km s$^{-1}$, cannot be explained by the Galactic rotation. We suggest that this feature represents an expanding structure driven by the stellar feedback of the progenitor of RCW~86.
The H{\sc i} expanding motion was also seen in the velocity channel distribution from $-45$ to $-25$ km s$^{-1}$ (see Appendix 
Figure \ref{fa1}) and the H{\sc i} line profile in Figure \ref{f05}b. We also show the $^{12}$CO($J$ = 2--1) contours in black. At $b > -2\fdg5$, the CO cloud has velocities higher (from $-40$ to $-35$ km s$^{-1}$) than the rest of the CO cloud, at $b < -2\fdg5$, for which velocities 
span from $-37$ to $-32$ km s$^{-1}$. 

\begin{figure*}
\begin{center}
\includegraphics[width=174mm,clip]{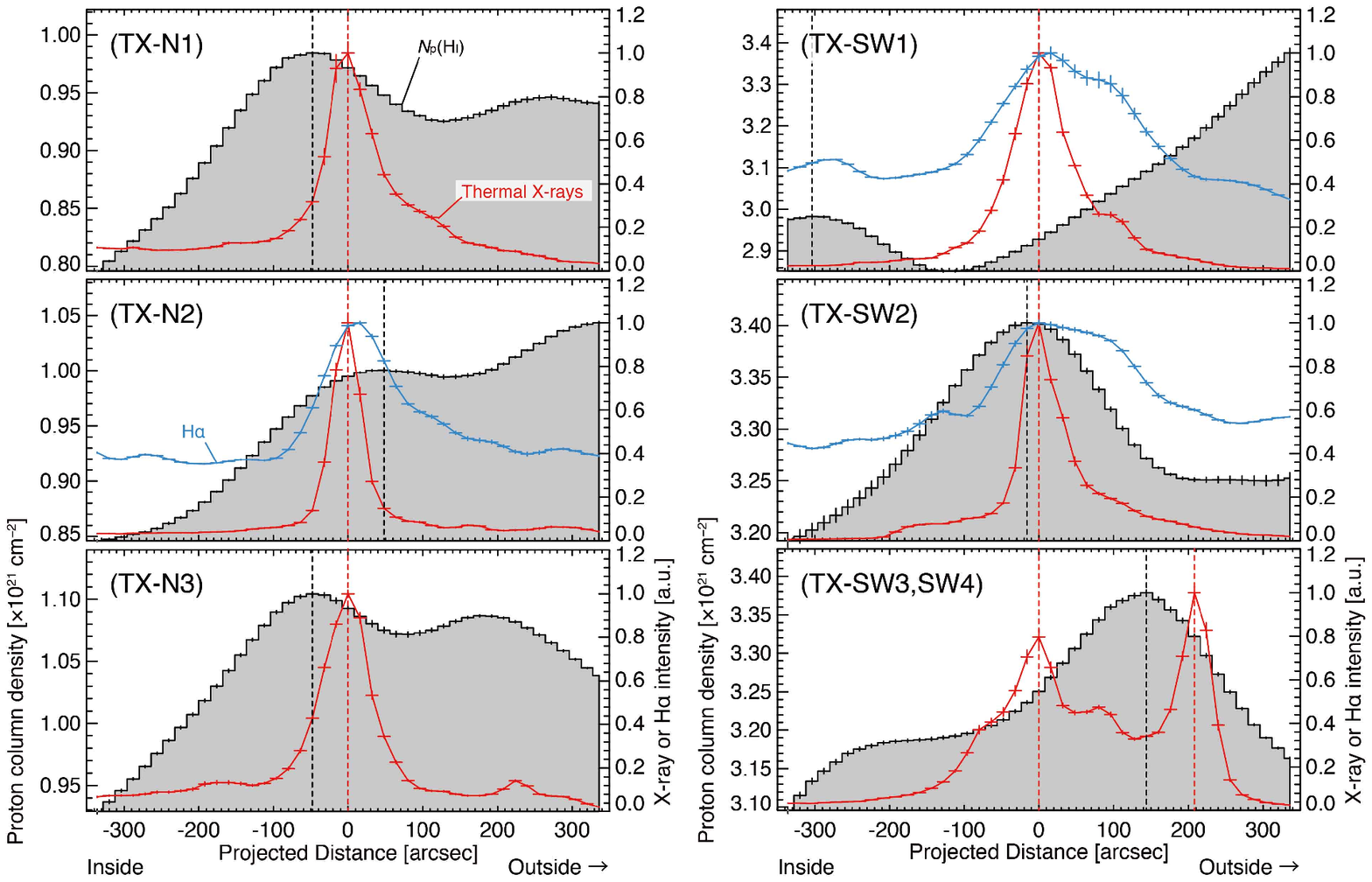}
\caption{Radial Profiles of the atomic proton column density $N_{\mathrm{p}}$(H{\sc i}) (gray filled areas), thermal X-rays (red), and H$\alpha$ emissions (light blue) for each rectangle region as shown by Figures \ref{f06}. Black, red, and light-blue dashed lines indicate the intensity peaks of $N_{\mathrm{p}}$(H{\sc i}), thermal X-rays, and H$\alpha$, respectively.}
\label{f07}
\end{center}
\end{figure*}%

The bright region of the H{\sc i} image is shifted toward the center of the SNR with a velocity increase from $-45$ to $-25$ km s$^{-1}$. We interpret that the H{\sc i} components of $-46$ km s$^{-1}$ and $-28$ km s$^{-1}$ correspond to the blue- and red-shifted sides of the expanding H{\sc i} wall, respectively. We note that the H{\sc i} intensity
of the red-shifted side is approximately twice as high as that of the blue-shifted side. If the emission is optically thin, the H{\sc i} intensity
corresponds to the mass.
By assuming this inhomogeneous gas distribution, the central velocity, $V_\mathrm{center}$, and expansion velocity, $\Delta V$, were estimated to be $V_\mathrm{center} \sim -35$ km s$^{-1}$ and $\Delta V \sim 7$--11 km s$^{-1}$, respectively. Here the central velocity corresponds to the kinematic distance of $\sim 2.4 \pm 0.2$ kpc adopting 
the Galactic rotation curve model of \citet{1993A&A...275...67B}. The error was derived using the uncertainty in the central velocity intrinsic to this method. 
The estimated distance is consistent with previous studies \citep[e.g.,][]{1969AJ.....74..879W,1996A&A...315..243R,2013MNRAS.435..910H,2016ApJ...819...98A}. The total mass and mean density of neutral atomic gas are estimated to be $\sim2\times10^4$ $M_{\odot}$ and $\sim70$ cm$^{-3}$, where the shell radius and thickness are assumed to be $\sim15$ pc and $\sim5$ pc, respectively \citep[c.f.,][]{2016arXiv160104461H}. H{\sc i} gas is generally considered to be optically thin (optical depth $\tau \ll 1$), having a column density, $N_\mathrm{p}$(H{\sc i})$'$ \citep[e.g.,][]{1990ARA&A..28..215D}:

\begin{eqnarray}
N_\mathrm{p}(\mathrm{H}{\textsc{i}})'= 1.823 \times 10^{18}  \int T_\mathrm{L}(V) dV (\mathrm{cm}^{-2}),  
\label{eq3}
\end{eqnarray}

where $T_\mathrm{L}(V)$ is the observed H{\sc i} brightness temperature in units of K. On the other hand, according to \cite{2015ApJ...798....6F}, 85 $\%$ of H{\sc i} gas is optically thick ($\tau \sim 0.5$--3) in the Milky Way, and the averaged column density is approximately 2--2.5 times higher than that derived on the optically thin assumption described by equation (\ref{eq3}). Subsequently, the authors established a more accurate relationship under consideration of the dust growth model (Fukui et al. 2017 in preparation). Therefore, we used the following relationship to calculate the ``true'' H{\sc i} column density, $N_\mathrm{p}(\mathrm{H}{\textsc{i}})$, instead of equation (\ref{eq3}):

\begin{eqnarray}
N_\mathrm{p}(\mathrm{H}{\textsc{i}}) = S \times N_\mathrm{p}(\mathrm{H}{\textsc{i}})' (\mathrm{cm}^{-2}),  
\label{eq4}
\end{eqnarray}

where $S$ is the conversion factor from $N_\mathrm{p}(\mathrm{H}{\textsc{i}})'$ to $N_\mathrm{p}(\mathrm{H}{\textsc{i}})$. In the region around RCW 86, the conversion factor, $S$, is estimated to be 2.3. Unless otherwise noted, we used equation (\ref{eq4}) and $S$ = 2.3 to calculate the H{\sc i} column density in this article. In the SNR RCW 86, $N_\mathrm{p}(\mathrm{H}{\textsc{i}})$ is accurately determined within $\sim8\%$, while the integrated intensity of H{\sc i} varies from 600 to 1,000 K km s$^{-1}$.

\subsection{Detailed Comparison with X-rays}
In order to establish a more detailed correspondence between the ISM and 
X-ray filaments in the velocity range from $-46$ km s$^{-1}$ to $-28$ km s$^{-1}$, we compare the integrated CO/H{\sc i} intensity map with the thermal and non-thermal X-rays. 

Figure \ref{f06} shows the intensity distribution of thermal X-rays, H{\sc i}, and CO. The H$\alpha$ contours with 500 dR or higher are also shown in the upper left of Figure \ref{f06}. We focused on the eastern, northern, and southwestern regions where thermal X-rays show filamentary distributions. In the eastern region the thermal X-ray filaments are distributed along with the CO $-$37 E cloud. The X-ray distribution cannot be interpreted by interstellar photoelectric absorption of the low-energy X-rays, because the thermal X-rays are not superposed onto the intensity peak of the CO cloud. We also found that the X-ray filament TX-E1 complex ($l$, $b$) $\sim$ ($315\fdg68$, $-2\fdg46$) is slightly aligned with the CO clumpy structure, while another filament TX-E2 complex, ($l$, $b$) $\sim$ ($315\fdg60$, $-2\fdg55$), is not much correlated with the CO cloud. This trend suggests that the degree of interaction between the SNR shocks and the CO cloud is different between the two regions. In the northern and southwestern regions, the distribution of the thermal X-rays shows a good spatial correlation with that of the H{\sc i} cavity wall at a scale of $\sim1$ pc, where the H{\sc i} intensity is significantly increased outwards from the SNR.

\begin{deluxetable*}{lccccccccc}
\tabletypesize{\scriptsize}
\tablecaption{Results of the Projected Profile towards the Thermal X-ray Peaks}
\tablewidth{0pt}
\tablehead{
&\multicolumn{3}{c}{Thermal X-rays} &  & \multicolumn{2}{c}{Separation}\\
\cline{2-4}\cline{6-7}
\multicolumn{1}{c}{Name}  & $l$ & $b$ & Peak Intensity & $<$$N_\mathrm{p}$(H{\sc i})$>$ & H{\sc i} Peak & H$\alpha$ Peak\\
& (deg) & (deg) & ($\times10^{-4}$ counts s$^{-1}$ pixel$^{-1}$) & ($\times10^{21}$ cm$^{-2}$) & (arcsec) & (arcsec) \\
\multicolumn{1}{c}{(1)} & (2) & (3) & (4) & (5) & (6) & (7)}
\startdata
TX-N1 & 315.58 & $-2.11$ & $1.30 \pm 0.07$ & $0.92 \pm 0.05$ & $-48$ & ------ \\
\hline
TX-N2 & 315.43 & $-2.06$ & $3.79 \pm 0.11$ & $0.96 \pm 0.07$ &$+32$  & 16 \\
\hline
TX-N3 & 315.23 & $-2.16$ & $3.94 \pm 0.07$ & $1.05 \pm 0.05$ &$-48$ & ------ \\
\hline
TX-SW1 & 315.14 & $-2.32$ & $12.6\phantom{0} \pm 0.3$\phantom{0}\phantom{0} & $3.03 \pm 0.15$ &$-304$\phantom{0} & 16 \\
\hline
TX-SW2 & 315.12 & $-2.47$ &  $20.6\phantom{0} \pm 0.8$\phantom{0}\phantom{0}& $3.29 \pm 0.06$ & $-16$ & \phantom{0}0 \\
\hline
TX-SW3 & 315.24 & $-2.51$ &  $\phantom{0}4.8\phantom{0} \pm 0.2\phantom{0}\phantom{0}$ & \multirow{2}{*}{$3.24 \pm 0.08$} & $+144$\phantom{0} & ------ \\
TX-SW4 & 315.22 & $-2.57$ & $\phantom{0}6.0\phantom{0} \pm 0.2\phantom{0}\phantom{0}$ & & $-68$ & ------
\enddata
\label{tab2}
\tablecomments{Col. (1): X-ray peak name. Cols. (2--4): Physical properties of the thermal X-rays. Cols. (2--3): Position of the X-ray peak. Col. (4): Peak intensity of the X-ray. Col. (5): Mean H{\sc i} column density, $<$$N_\mathrm{p}$(H{\sc i})$>$,  within each region shown by Figure \ref{f06}. Col. (6): Separation between each intensity peak of the X-ray and H{\sc i} emissions. Col. (7): Separation between each intensity peak of the X-ray and H$\alpha$ emissions.}
\end{deluxetable*}

Figure \ref{f07} shows the radial profiles of the proton column density $N_\mathrm{p}(\mathrm{H}{\textsc{i}})$ (gray filled areas) and the thermal X-ray intensity (red) for each region of dashed rectangles, perpendicular to the shell as shown in Figure \ref{f06}. Each region has a $670'' \times 160''$ size corresponding to 8 pc $\times$ 2 pc, and is centered on the X-ray filament (see Table \ref{tab2}). We defined the origin of the radial profile as the position of the maximum X-ray intensity in the projected distance. Positive and negative values correspond to the outer and inner sides of the SNR shell, respectively. The regions are selected every $\sim3$ pc relative to the azimuthal direction, and all of them cross local X-ray peaks. We also added the H$\alpha$ distribution (light blue) in the TX-N2, TX-SW1, and TX-SW2 regions, which have H$\alpha$ fluxes of 500 dR or higher. The intensity scales of the thermal X-rays and H$\alpha$ are normalized by their maximum values, and the positions of the intensity peaks are indicated by the vertical dashed lines. We find that the positions of the H{\sc i} intensity peaks correspond well with those of X-rays and H$\alpha$, except in TX-SW1. In order to evaluate quantitatively this trend, we estimated the accurate values of intensity peaks on the radial profiles. Table \ref{tab2} shows a summary of the radial profile towards each thermal X-ray peak. We defined the separation from the X-ray intensity peak to the H{\sc i} or H$\alpha$ intensity peaks in the radial distribution, in such a way that a positive (negative) value 
implies that the X-ray peak lies to the left (right) of the other peaks in the diagram. The separations between the thermal X-ray peak and the H{\sc i}/H$\alpha$ peaks are smaller than the beam size of the H{\sc i} data, $\sim156$ arcsec, except for the H{\sc i} peak of TX-SW1. 

\begin{figure}
\begin{center}
\includegraphics[width=87mm,clip]{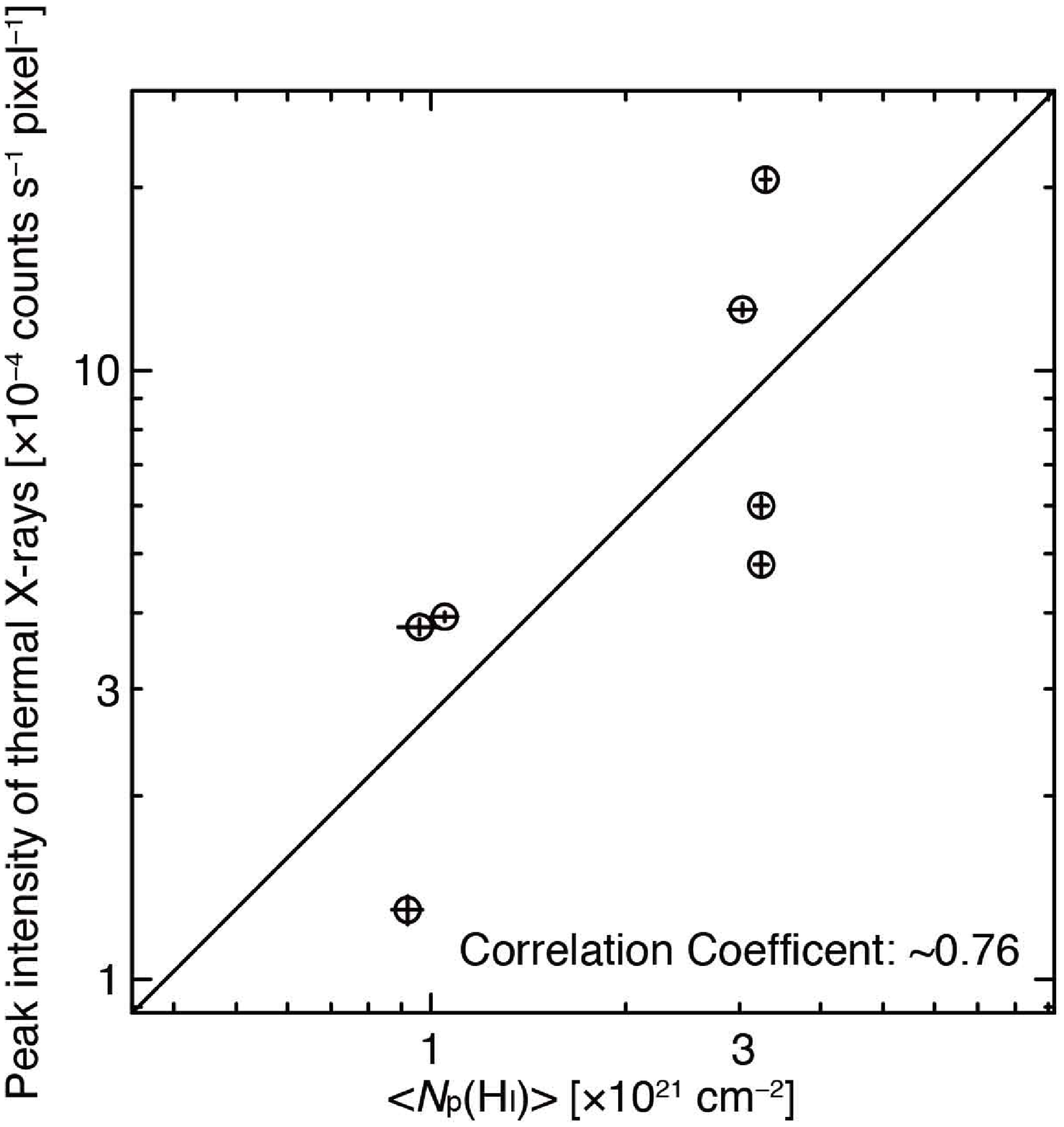}
\caption{Correlation plot between the averaged H{\sc i} column density $<$$N_\mathrm{p}$(H{\sc i})$>$, and the peak intensity of thermal X-rays. The error bars are also indicated. The solid line shows the linear regression of the double logarithm plot applying a least squares fit, where the correlation coefficient is $\sim 0.76$.}
\label{f08}
\end{center}
\end{figure}

We also estimated the mean H{\sc i} column density $<$$N_\mathrm{p}(\mathrm{H}{\textsc{i}})$$>$ within each rectangle. The double-logarithm plot in Figure \ref{f08} shows the correlation 
between $<$$N_\mathrm{p}(\mathrm{H}{\textsc{i}})$$>$ and the 
peak intensity of the thermal X-rays. The solid line shows the linear regression
by least-squares fitting, with a correlation coefficient of $\sim0.76$. We conclude that the thermal X-ray intensity increases following roughly a power-law dependence with the column density of neutral atomic gas at a pc scale.

\begin{figure*}
\begin{center}
\includegraphics[width=166mm,clip]{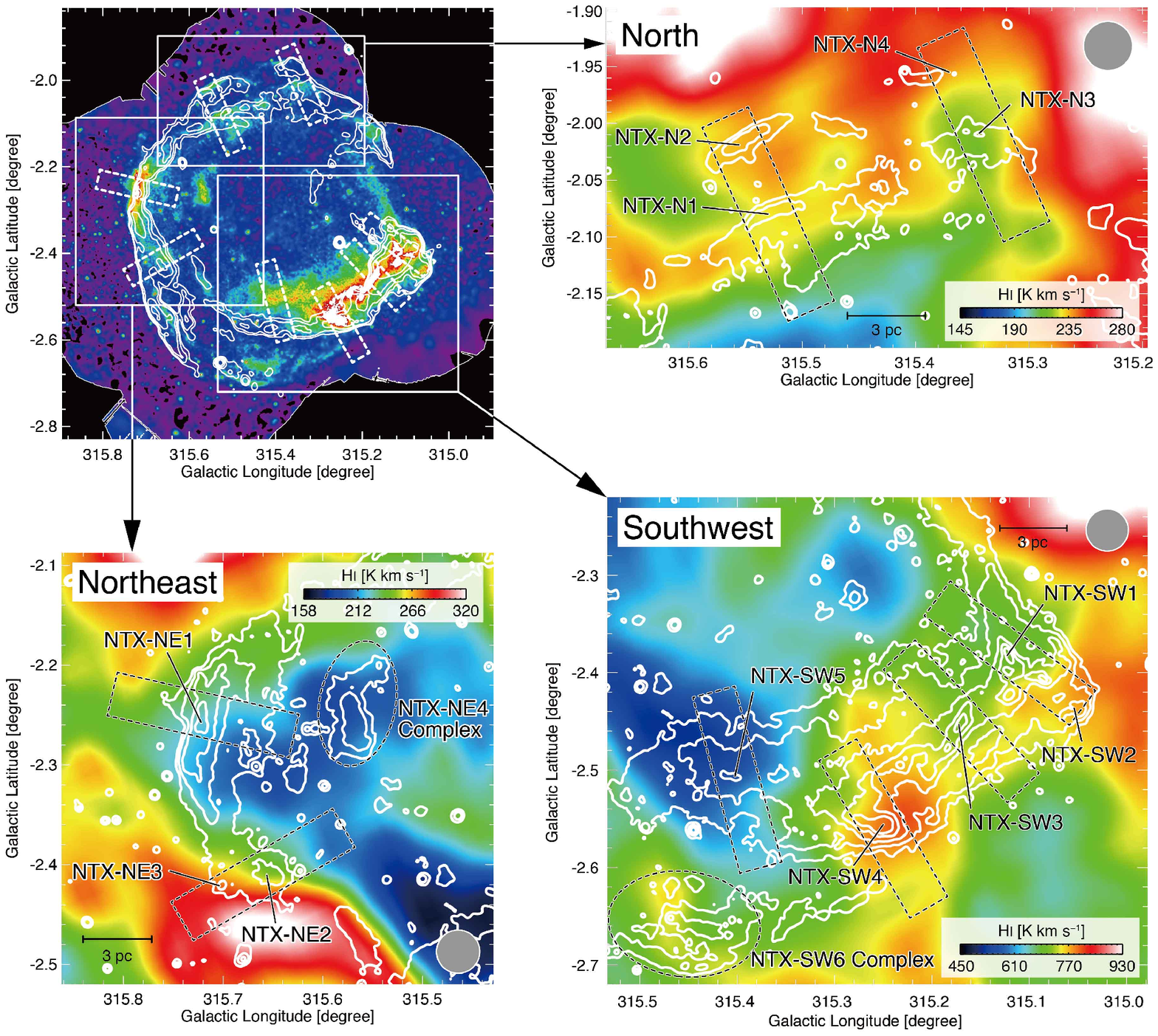}
\caption{Top left: $XMM$-$Newton$ X-ray image in the energy band of 2.0--4.5 keV overlaid with MOST radio continuum contours as shown in Figure \ref{f01}. Other three panels: distributions of H{\sc i} obtained with ATCA $\&$ Parkes (rainbow scale) superposed with the $XMM$-$Newton$ X-ray contours in the energy band of 2.0--4.5 keV. The three regions toward the north, the northeast, and the southwest are shown with inserts in the X-ray image on the top left panel. The velocity range spans from $-32$ to $-29$ km s$^{-1}$ in the north; from $-38$ to $-33$ km s$^{-1}$ in the northeast; and from $-42$ to $-28$ km s$^{-1}$ in the southwest. The contour levels of the X-rays are 8.0$\times$10$^{-8}$, 1.7$\times$10$^{-5}$, 4.4$\times$10$^{-5}$, 8.8$\times$10$^{-5}$, and 1.5$\times$10$^{-4}$ photons s$^{-1}$ pixel$^{-1}$.}
\label{f09}
\end{center}
\end{figure*}

Figure \ref{f09} shows the intensity distribution of the non-thermal X-rays and 
H{\sc i}. We focused on the northern, southwestern, and northeastern regions, 
where non-thermal X-rays are prominent. In the northern and southwestern regions, the non-thermal X-ray filaments are spatially well correlated with the H{\sc i} bright-rim at a pc scale, as well as the thermal X-rays in Figure \ref{f07}. In contrast, the X-ray peaks NTX-NE1 and NTX-SW5 are located inside the H{\sc i} bright wall, while the shape of NTX-NE1 filament slightly matches the H{\sc i} distribution. In addition to this, we also find 
that the non-thermal X-ray complex of NTX-NE4, ($l$, $b$) $\sim$ ($315\fdg57$, $-2\fdg24$), is located inwards with respect to NTX-NE1, while the NTX-SW6 complex, ($l$, $b$) $\sim$ ($315\fdg44$, $-2\fdg66$), lies outwards from NTX-SW5.

\begin{figure*}
\begin{center}
\includegraphics[width=174mm,clip]{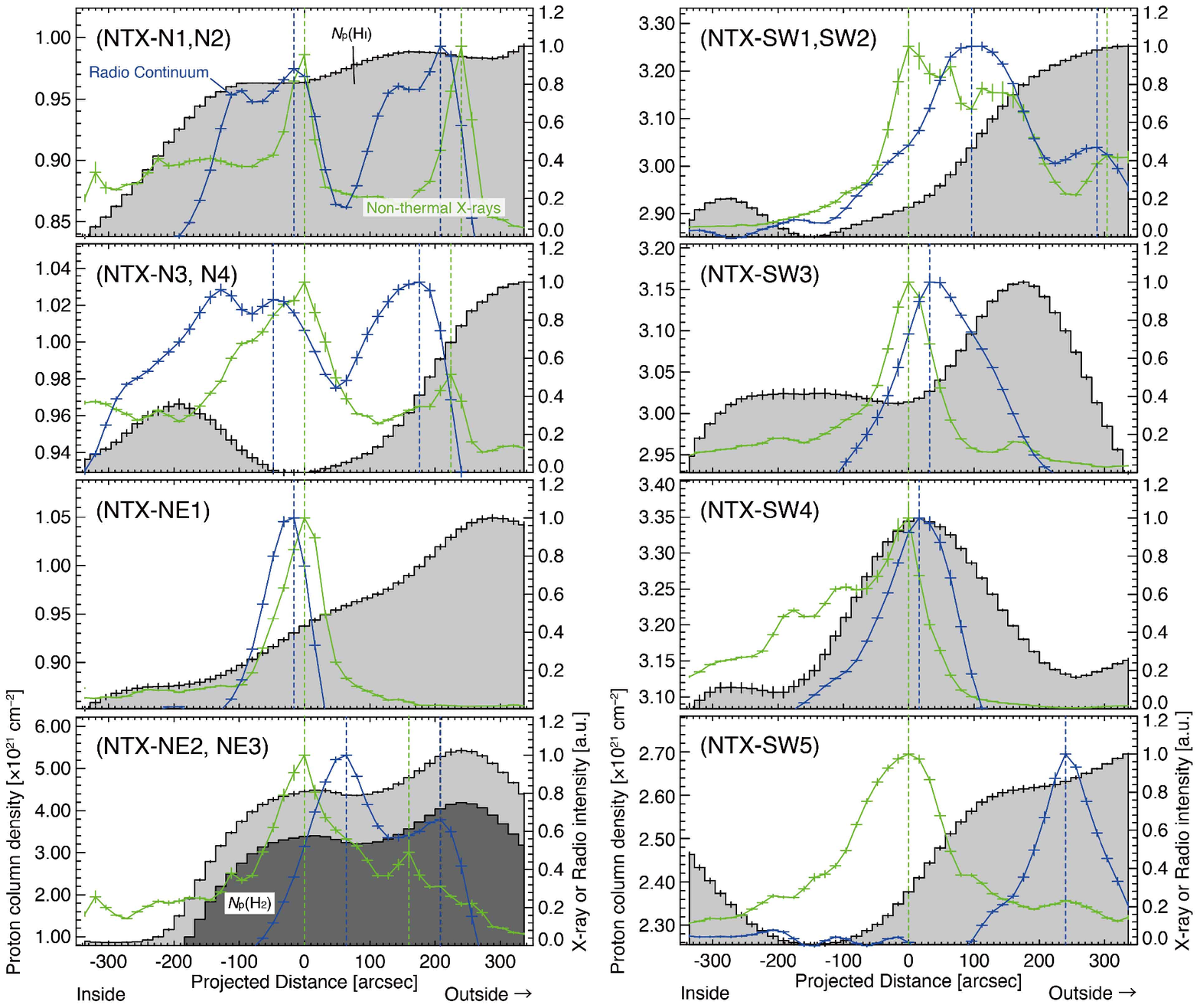}
\caption{Radial profiles of proton column density (black), non-thermal X-rays (green), and radio continuum (blue) for each rectangle region, as shown by Figures \ref{f09}. Gray and black filled areas represent the atomic proton column density, $N_{\mathrm{p}}$(H{\sc i}), and the molecular proton column density, $N_{\mathrm{p}}$(H$_2$), respectively. Green and blue dashed lines indicate the intensity peaks of non-thermal X-rays and radio continuum radiation, respectively.}
\label{f10}
\end{center}
\end{figure*}

We analyzed each non-thermal X-ray peak in a manner similar to the thermal X-ray case. Figure \ref{f10} shows the radial profiles of the proton column density, $N_\mathrm{p}(\mathrm{H}{\textsc{i}})$ (gray filled areas), and non-thermal X-ray intensity (green) for each rectangle, as shown by Figure \ref{f09}. Here, we considered the column density of the molecular hydrogen, $N(\mathrm{H}_2)$, toward the NEX-NE2, -NE3 region, where there is a significant amount of molecular mass. Finally, we estimated the total proton column density, $N_\mathrm{p}$(H$_2 + $H{\sc i}), by

\begin{eqnarray}
N_\mathrm{p}(\mathrm{H}_2 + \mathrm{H}{\textsc{i}}) = 2 \times N(\mathrm{H}_2) +  N_\mathrm{p}(\mathrm{H}{\textsc{i}}),  
\label{eq5}
\end{eqnarray}

\begin{figure}
\begin{center}
\includegraphics[width=87mm,clip]{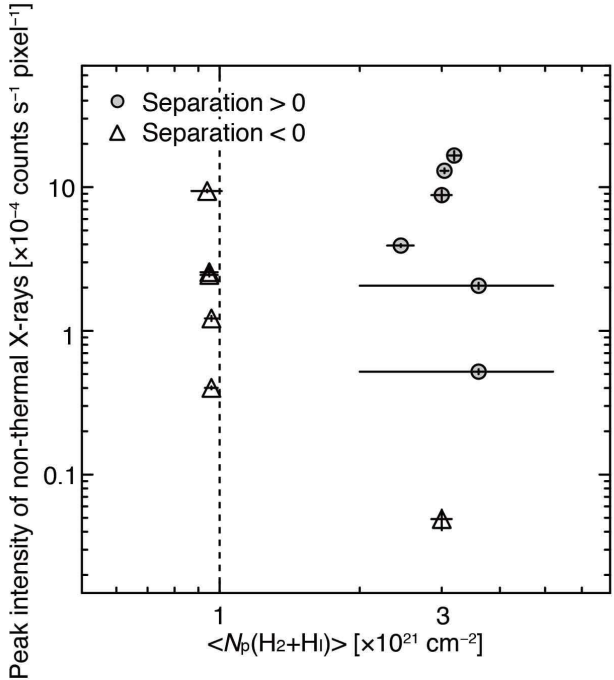}
\caption{Correlation plot between the averaged proton column density $<$$N_\mathrm{p}$(H$_2 + $H{\sc i})$>$, and the peak intensity of non-thermal X-rays. 
Positive and negative separations (see text) are represented by circles and triangles, respectively, where the separation is defined as the distance between the non-thermal X-rays and radio continuum intensity peaks. The vertical dashed line indicates $<$$N_\mathrm{p}$(H$_2+$H{\sc i})$>$ = $1 \times10^{21}$ cm$^{-2}$.}
\label{f11}
\end{center}
\end{figure}

We also added the radio continuum distribution (blue) in all the regions. The difference in the distributions of the non-thermal X-rays and the radio continuum indicate the energy difference of the CR electrons. The X-ray and radio peaks are located around the H{\sc i} intensity peaks. We did not find a specific trend among them, such as the correlation between X-rays and H{\sc i}. The position of the intensity peaks is represented by the vertical dashed lines of both the X-ray and radio peaks. We note that the relative radial positions between the X-ray and radio peaks show significant offsets from each other. Specifically, the non-thermal X-ray intensity peaks NTX-N1--N4, NTX-NE1, and NTX-SW2 are positioned farther outside than the nearest radio peaks, while the NTX-NE2--3, NTX-SW1, and NTX-SW3--5 peaks are located farther inside than the nearest radio peaks. Table \ref{tab3} shows the trend quantitatively. The positive (negative) values of the separation correspond to the case in which the non-thermal X-ray intensity peaks are positioned inwards (outwards) from the nearest radio peaks. Most of the separations are larger than half of the beam size for both the radio and X-rays ($\Delta \theta \sim 20$ arcsec). Considering the extended radial distribution of the radio peaks, all the separations are regarded to be significant. Figure \ref{f11} shows a logarithmic plot of the correlation between the averaged total proton column density $<$$N_\mathrm{p}(\mathrm{H}_2 + \mathrm{H}{\textsc{i}})$$>$ and the peak intensity of the non-thermal X-rays. Filled circles represent positive separations, while open triangles are used to plot negative ones. The vertical dashed line indicates $<$$N_\mathrm{p}(\mathrm{H}_2 + \mathrm{H}{\textsc{i}})$$>$ = $1 \times 10^{21}$ cm$^{-2}$. In contrast to the thermal X-ray case, there is no significant correlation between the non-thermal X-rays and $<$$N_\mathrm{p}(\mathrm{H}_2 + \mathrm{H}{\textsc{i}})$$>$: if a least-square fitting is attempted to the double logarithm plot, the correlation coefficient turns out to be $\sim0.03$. We find that, with the only exception of NTX-SW2, the negative separations are clustered within the low-density region ($\sim1 \times 10^{21}$ cm$^{-2}$), whereas the positive separations are only located in regions of density higher than $\sim3 \times 10^{21}$ cm$^{-2}$. 

\begin{deluxetable*}{lcccccccc}[h]
\tabletypesize{\scriptsize}
\tablecaption{Results of the Projected Profile towards the Non-Thermal X-ray Peaks}
\tablewidth{0pt}
\tablehead{
&\multicolumn{3}{c}{Non-Thermal X-rays} & & Separation\\
\cline{2-4}\cline{6-6}
\multicolumn{1}{c}{Name}  & $l$ & $b$ & Peak Intensity & $<$$N_\mathrm{p}$(H$_2+$H{\sc i})$>$ & Radio Peak \\
& (deg) & (deg) & ($\times10^{-5}$ counts s$^{-1}$ pixel$^{-1}$) & ($\times10^{21}$ cm$^{-2}$) & (arcsec) \\
\multicolumn{1}{c}{(1)} & (2) & (3) & (4) & (5) & (6)}
\startdata
NTX-N1 & 315.53 & $-2.08$ & $2.56\phantom{0} \pm 0.11\phantom{0}$ & \multirow{2}{*}{$0.95 \pm 0.04$} & $-16$\\
NTX-N2 & 315.56 & $-2.02$ & $2.45 \phantom{0} \pm 0.06\phantom{0}$ & & $-32$\\
\hline
NTX-N3 & 315.34 & $-2.01$ & $1.22 \phantom{0} \pm 0.05\phantom{0}$ & \multirow{2}{*}{$0.96 \pm 0.03$} & $-48$\\
NTX-N4 & 315.37 & $-1.96$ & $0.401 \pm 0.009$ & & $-48$\\
\hline
NTX-NE1 & 315.72 & $-2.25$ & $9.4\phantom{0}\phantom{0} \pm 0.2\phantom{0}\phantom{0}$ & $0.94 \pm 0.07$ & $-16$\\
\hline
NTX-NE2 & 315.66 & $-2.41$ & $2.06\phantom{0} \pm 0.10\phantom{0}$ & \multirow{2}{*}{$3.6\phantom{0} \pm 1.6\phantom{0}$} & $+64$\\
NTX-NE3 & 315.70 & $-2.42$ & $0.52\phantom{0} \pm 0.02\phantom{0}$ & & $+48$\\
\hline
NTX-SW1 & 315.12 & $-2.38$ & $8.8 \phantom{0}\phantom{0}\pm 0.7\phantom{0}\phantom{0}$ & \multirow{2}{*}{$3.00 \pm 0.15$} & $+96$\\
NTX-SW2 & 315.05 & $-2.44$ & $0.490 \pm 0.008$ & & $-16$\\
\hline
NTX-SW3 & 315.17 & $-2.45$ & $13.0\phantom{0}\phantom{0} \pm 0.5$\phantom{0}\phantom{0}\phantom{0}& $3.04\pm 0.06$ & $+32$\\
\hline
NTX-SW4 & 315.25 & $-2.56$ & $16.6\phantom{0}\phantom{0} \pm 1.2$\phantom{0}\phantom{0}\phantom{0} & $3.19 \pm 0.09$ & $+16$\\
\hline
NTX-SW5 & 315.40 & $-2.51$ & $3.92\phantom{0} \pm 0.09\phantom{0}$& $2.45\pm 0.16$ & $+240$\phantom{0}
\enddata
\label{tab3}
\tablecomments{Col. (1): X-ray peak name. Cols. (2--4): Physical properties of the non-thermal X-rays. Cols. (2--3): Position of the X-ray peak. Col. (4): Peak intensity of the X-ray. Col. (5): Mean proton column density $<$$N_\mathrm{p}$(H$_2+$H{\sc i})$>$ within each region shown by Figure \ref{f09}. Col. (6): Separation between each intensity peak of the X-ray and radio continua.}
\end{deluxetable*}

\section{DISCUSSION}
\subsection{Progenitor System of RCW 86}
There have been considerable debates on the progenitor system (CC or Type Ia) of RCW86 since its discovery \citep{1969AJ.....74..879W,1989ApJ...337..399C,1992A&A...264..654K,1997A&A...328..628V,2000PASJ...52.1157B,2011PASJ...63S.837Y,2011ApJ...741...96W,2014MNRAS.441.3040B}. Recent multi-wavelength observations as well as theoretical studies in the last several years reveal the progenitor is a Type Ia SN. \cite{2007PASJ...59S.171U} and \cite{2008PASJ...60S.123Y} found that the abundant Fe ejecta and the absence of rich O ejecta are consistent with a Type Ia SNR. \cite{2011ApJ...741...96W} argued that the H$\alpha$ filamentary distributions are created by the interaction between the SNR shocks and the neutral gas and suggested that the interaction between the SNR shock and the ambient gas was weak \citep[e.g.,][]{1980ApJ...235..186C,1997AJ....114.2664S}. This is consistent with the accretion wind by a Type Ia progenitor 
\citep[e.g.,][]{1996ApJ...470L..97H,2007ApJ...663.1269N}.  A central compact stellar remnant like a neutron star or a pulsar wind nebula is not yet detected, again favoring a Type Ia \citep[e.g.,][]{2004ApJS..153..269K}. \cite{2011ApJ...741...96W} calculated an off-center explosion by 
using a 2D hydrodynamic model, applying parameters given by the Type Ia progenitor accretion wind model proposed by \cite{2007ApJ...662..472B}. The authors showed that the size of the SNR, shock-velocity, and post-shock gas density are well reproduced, concluding that RCW 86 is Type Ia explosion in an accretion-wind bubble. 
In this section we discuss the progenitor system of RCW 86 based on the results of the associated interstellar gas. 

\begin{figure*}
\begin{center}
\includegraphics[width=166mm,clip]{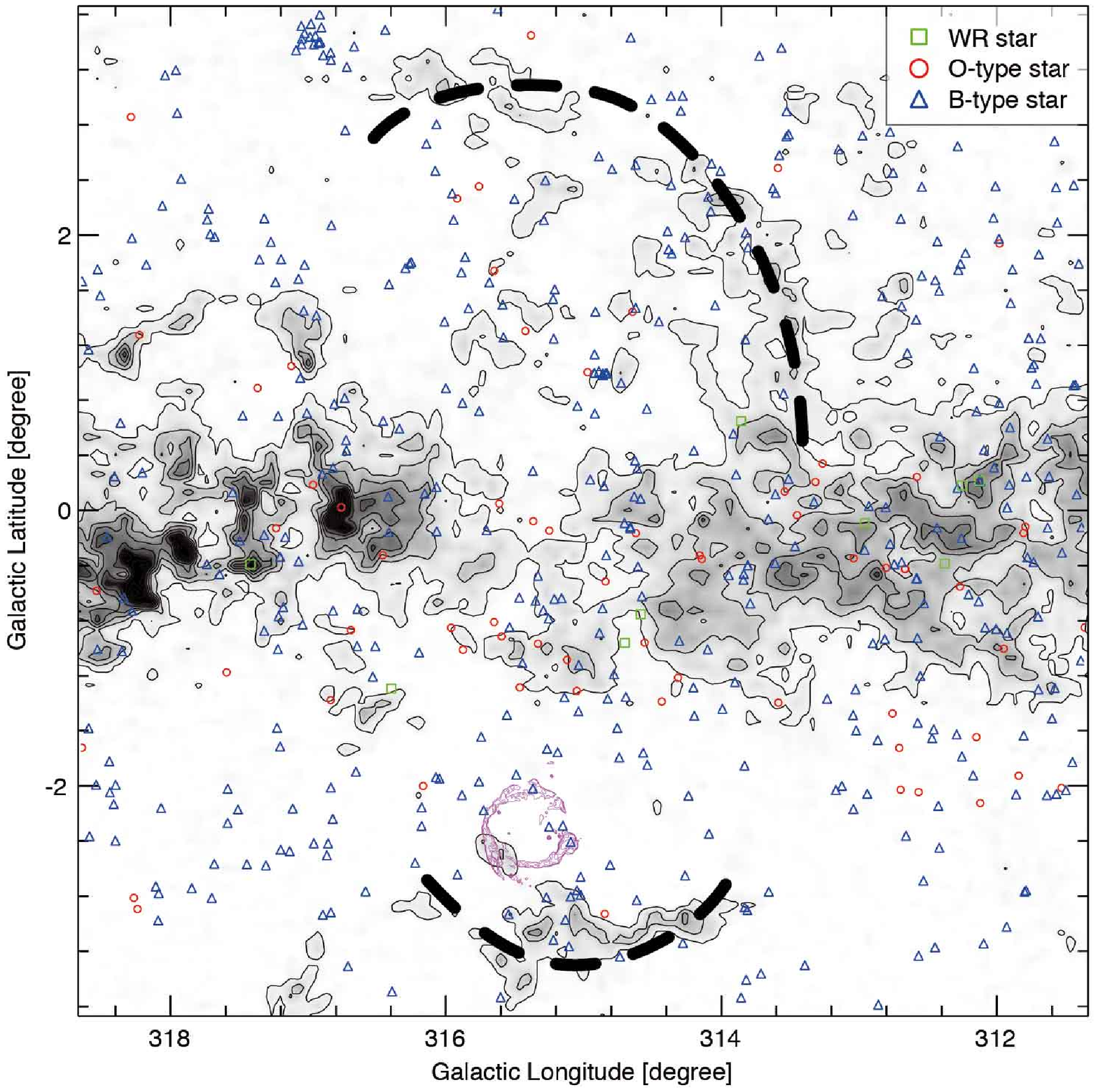}
\caption{Large-scale integrated intensity map of the $^{12}$CO($J$ = 1--0) toward the RCW 86 taken with NANTEN (Matsunaga et al. 2001). The velocity range is from $-40$ to $-30$ km s$^{-1}$. The lowest contour level and contour interval of CO are 10 K km s$^{-1}$ and 4 K km s$^{-1}$, respectively. Magenta contours indicate the MOST radio continuum at a frequency of 843 MHz as shown in Figure \ref{f01}. The dashed lines represent the boundary of the CO supershell identified by \citet{2001PASJ...53.1003M}.}
\label{f12}
\end{center}
\end{figure*}%

First, we shall argue that the H{\sc i}/CO expanding structure is inconsistent with the acceleration by only the accretion wind of the progenitor. The expansion velocity of H{\sc i} $\sim7$ km s$^{-1}$ in the red-shifted (gas-rich) side and the total H{\sc i} mass $2\times10^4$ $M_{\odot}$ lead to a momentum of $\sim3 \times 10^{38}$ N$\cdot$s for the H{\sc i} shell. Figure \ref{f05}a shows that the CO peak velocity is shifted by $\sim$3 km towards the interior of the SNR. The momentum of this shifted component is only 5 $\%$ of the whole CO $-37$ E cloud and the
CO kinetic energy is negligible as compared to 
the H{\sc i} kinetic energy. \cite{1996ApJ...470L..97H} showed that the accretion wind of a Type Ia progenitor has a typical duration of $3 \times 10^5$ yr, where the wind mass and wind velocity are $\sim10^{-6} M_{\odot}$ yr$^{-1}$ and $\sim1,000$ km s$^{-1}$, respectively. This means that the momentum released by the accretion wind amounts to $\sim6 \times 10^{35}$ N$\cdot$s, which is quite small to explain the observed momentum of the H{\sc i} shell. Moreover, it is difficult to explain the shell formation in terms of the SN shock waves. RCW 86 has an age of $\sim1,800$ yr and the duration of the shock interaction with the ambient medium is too short to transfer the momentum significantly. This is consistent with the absence of wing like emission spaning more than 10 km s$^{-1}$ in the CO spectra \citep[e.g.,][]{1998ApJ...505..286S,2013ApJ...768..179Y} and the fact that only a thin surface of CO gas is heated by shock interaction (see also Figure \ref{f04}). The short duration of the interaction is also suggested from the X-ray spectroscopy. \cite{1997A&A...328..628V} showed that the thermal plasma in RCW 86 is dramatically deviated from the thermal equilibrium and noted that there is a spot of extremely short ionization timescale in the SNR. In particular, the time elapsed since the Fe ejecta was heated by the reverse SNR shock is estimated to be $<$ 380 yr, a fourth of the SNR age \citep[e.g.,][]{2008PASJ...60S.123Y}. We suggest that the H{\sc i}/CO expanding structure could be formed by the stellar winds from nearby OB stars \citep[e.g.,][]{1969AJ.....74..879W}. Figure \ref{f12} shows that RCW 86 is at the inner edge of a molecular supershell created by multiple supernova remnants in the Galactic plane. The expansion in RCW 86 may originate in the supershell.

\begin{figure*}
\begin{center}
\includegraphics[width=180mm,clip]{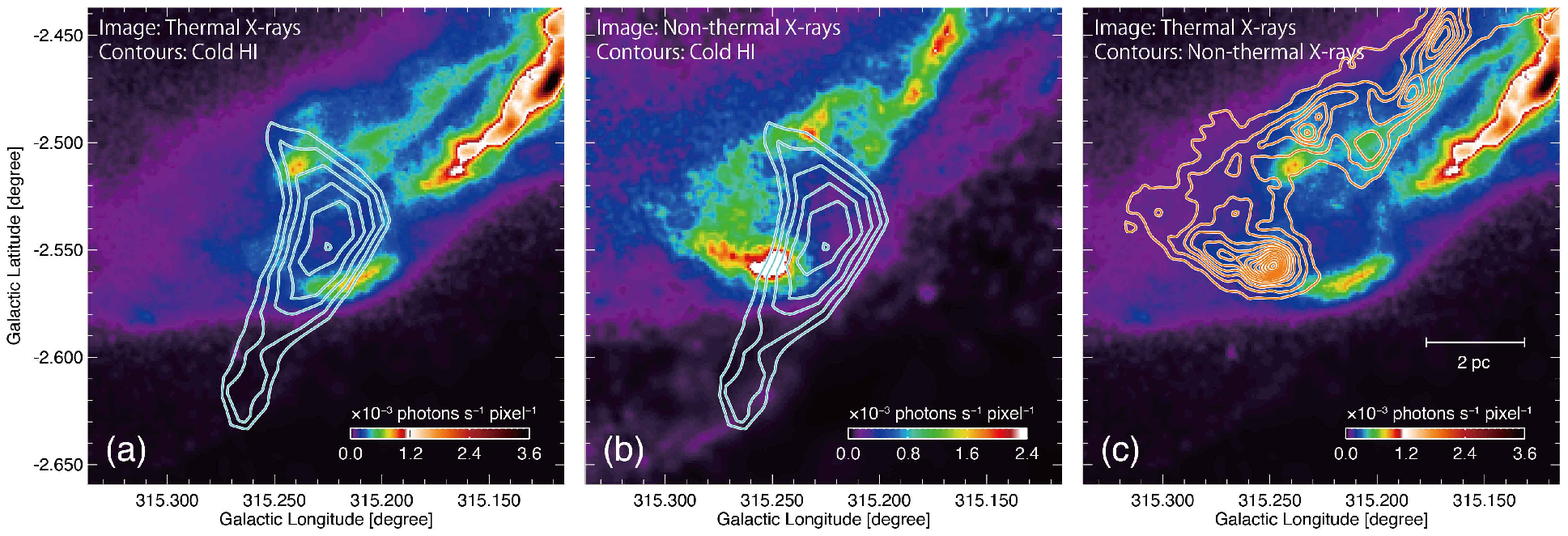}
\caption{(a) Thermal X-ray image of RCW 86 in the energy range from 0.5 to 1.0 keV superposed with H{\sc i} intensity contours (light blue) toward the southwest of the SNR. The color scale is linear, in units of $10^3$ photons s$^{-1}$ pixel$^{-1}$. The H{\sc i} velocity is $V_\mathrm{LSR}$ = $-$35 km s$^{-1}$. The lowest contour level and contour interval of H{\sc i} are 54 K ($\sim 72 \sigma$) and 0.75 K ($\sim 1 \sigma$), respectively. (b) Non-thermal X-ray image in the energy range from 2.0 to 4.5 keV, superposed with H{\sc i} intensity contours. The color scale, unit, and contour levels are the same as in the left panel. (c) The same thermal X-ray image as Figure \ref{f13}a, superposed with the non-thermal X-ray contours (orange). The lowest contour level and contour interval of the non-thermal X-rays are $5 \times 10^{-5} $ photons s$^{-1}$ pixel$^{-1}$ and $2.5 \times 10^{-5} $ photons s$^{-1}$ pixel$^{-1}$, respectively.}
\label{f13}
\end{center}
\end{figure*}%

Comparison with other CC SNRs of similar ages and properties reinforces that RCW 86 is a Type Ia SNR with a low-velocity wind. Because of their similarities,
RX J1713.7$-$3946, a CC shell SNR with an age of $\sim1,600$ years, emitting bright TeV $\gamma$-ray and non-thermal X-rays \citep{2003PASJ...55L..61F,2004A&A...427..199C, 2005ApJ...631..947M}, is the best target to compare with RCW 86. In RX J1713.7$-$3946, molecular clouds of $n$ $> 10^4$ cm$^{-3}$ remained without being swept 
up by the SNR shock wave, whereas the intercloud and diffuse H{\sc i} gas were evacuated by the strong stellar winds from the massive progenitor \citep[e.g.,][]{2003PASJ...55L..61F,2005ApJ...631..947M,2010ApJ...724...59S,2013ApJ...778...59S,2013PASA...30...55M}. As a result, in RX J1713.7$-$3946 we do not see the H{\sc i} envelope of CO clouds and strong thermal X-rays are not detected \citep[e.g.,][]{2008PASJ...60S.131T,2012ApJ...746...82F,2015ApJ...799..175S}. In RCW 86, instead, we see diffuse H{\sc i} toward the CO clouds. In particular, we can clearly see an H{\sc i} envelope around CO $-37$ E (see Figure \ref{f02}). This suggests that the progenitor had a weaker wind than the massive progenitor of the CC SNR RX J1713.7$-$3946 and is consistent with the accretion wind hypothesis. The CC scenario with an early B-star, nevertheless, cannot be ruled out. The thermal X-rays observed over the whole RCW 86 suggest that a large amount of H{\sc i} gas is distributed inside the shell. 

These thermal X-rays are produced by the interaction of shock waves with the 
preexistent neutral and ionized gas,
even though observational evidence for the interaction 
was not obtained \citep[e.g.,][]{2002ApJ...581.1116R,2011PASJ...63S.837Y}. Figure \ref{f07} shows that the thermal X-ray peaks coincide with the H{\sc i} peaks, indicating that the shock waves have collided into the H{\sc i} cavity-wall and radiate the thermal X-rays. Only in TX-S1 the H{\sc i} does not coincide with the X-ray peak. This difference is explained if we assume that the H{\sc i} associated with TX-S1 is already ionized, because the X-rays peak at the H$\alpha$ peak and the electron density is estimated to be $\sim100$ cm$^{-3}$ in the South \citep{1981ApJ...243..814R}. In addition, the intensity of the thermal X-rays increases with the neutral gas density, suggesting that the gas is thermalized by the shock passage. Detailed spectrum analysis of X-rays comparable to the interstellar distribution will reveal a better correlation between the proton column density and thermal X-ray intensity. Based on the considerations above, we conclude that RCW 86 is the remnant of a Type Ia explosion in
a wind-bubble and state that a thorough investigation of the neutral gas is an important tool to investigate the progenitor system and the origin of the thermal X-rays. 

\subsection{Efficient CR Acceleration}
\cite{2015ApJ...799..175S} argued that in RX J1713.7$-$3946, the efficient CR electron acceleration up to $\sim10$ TeV currently at
work has a tight physical connection with the ambient ISM. The
authors showed that the distribution of the photon index $\Gamma$ of the non-thermal X-rays, synchrotron X-rays, and both the gas-rich and -poor regions is small, with $\Gamma < 2.4$, and suggested that these regions correspond to the sites of high roll-off energy of the synchrotron emission. If the synchrotron cooling is efficient, the roll-off energy $\varepsilon_0$ of the synchrotron photons is given by the following equation \citep{2007A&A...465..695Z}:

\begin{eqnarray}
\varepsilon_0 = 0.55 \times (v_\mathrm{sh}\:/\:3000 \;\mathrm{km\; s^{-1}})^2 \:\eta^{-1} \;\mathrm{(keV)},
\label{eq6}
\end{eqnarray}

where $v_\mathrm{sh}$ is the shock velocity, and $\eta$ = $B^2$/$\delta B^2$ ($> 1$) the degree of magnetic field fluctuations (the gyro-factor). The case of $\eta = 1$ is called the Bohm diffusion limit and indicates the limit of the maximum magnetic turbulence. In the gas-rich/clumpy site, the shock-cloud interaction amplifies the turbulent magnetic field around dense gas clumps and the synchrotron X-rays are enhanced \citep[e.g.,][]{2012ApJ...744...71I}. As a consequence, $\eta \sim 1$, hence increasing $\varepsilon_0$. On the other hand, in the gas-poor/diffuse sites the shock waves are not decelerated, and the high $v_\mathrm{sh}$ results also in this case in a high $\varepsilon_0$ and, accordingly, in an enhancement in the X-rays emission. Sano et al. argued that the ambient conditions of the neutral ISM play a role in increasing the roll-off energy in the CR acceleration and the non-thermal X-rays. In what follows, we discuss the properties of the non-thermal X-rays in RCW 86 by comparing the X-ray properties in RX J1713.7$-$3946. The most intense non-thermal X-ray filaments in RCW 86 (Figure \ref{f01}) are seen at the NE and SW. The average proton column density $<$$N_\mathrm{p}(\mathrm{H}_2 + \mathrm{H}{\textsc{i}})$$>$ is 0.94 $\pm$ 0.07 $ \times 10^{21}$ cm$^{-2}$ in NTX-NE1 and 3.19 $\pm$ 0.09 $\times$ 10$^{21}$ cm$^{-2}$ in NTX-SW4, i.e., they differ by a factor of three. In RCW 86 the gas-rich/clumpy region corresponds to the SW, and the gas-poor/diffuse region, to the NE. At the SW, however, the H{\sc i} cloud does not have a CO clumpy
counterpart. Unlike RX~J1713.7$-$3946, there were no cold H{\sc i} clumps detected as self-absorption features. Figure \ref{f13} shows the X-rays and the H{\sc i} clump in SW. We see the thermal/non-thermal X-rays are enhanced around the cold H{\sc i} clump. This suggests that shock-cloud interaction with the cold H{\sc i} clump amplifies turbulence and magnetic field, causing the rim-brightened non-thermal X-rays. A similar enhancement of the thermal X-rays is seen. This indicates that the shock waves are heating up the surface of the cold H{\sc i} clump. The H{\sc i} peak brightness temperature of the clump is low, $\sim58$ K, suggesting that the clump has density $\sim150$ cm$^{-3}$ \citep{2014ApJ...796...59F,2015ApJ...798....6F}. The same complementary spatial distribution between cold H{\sc i} and
X-rays due to shock-cloud interaction is also observed in RX~J1713.7$-$3946 \citep[see Figure 4 of ][]{2013ApJ...778...59S}. It is also expected that the synchrotron X-ray flux will vary within a scale of several years due to the strong magnetic field \citep[e.g.,][]{2007Natur.449..576U}.

\begin{figure*}
\begin{center}
\includegraphics[width=180mm,clip]{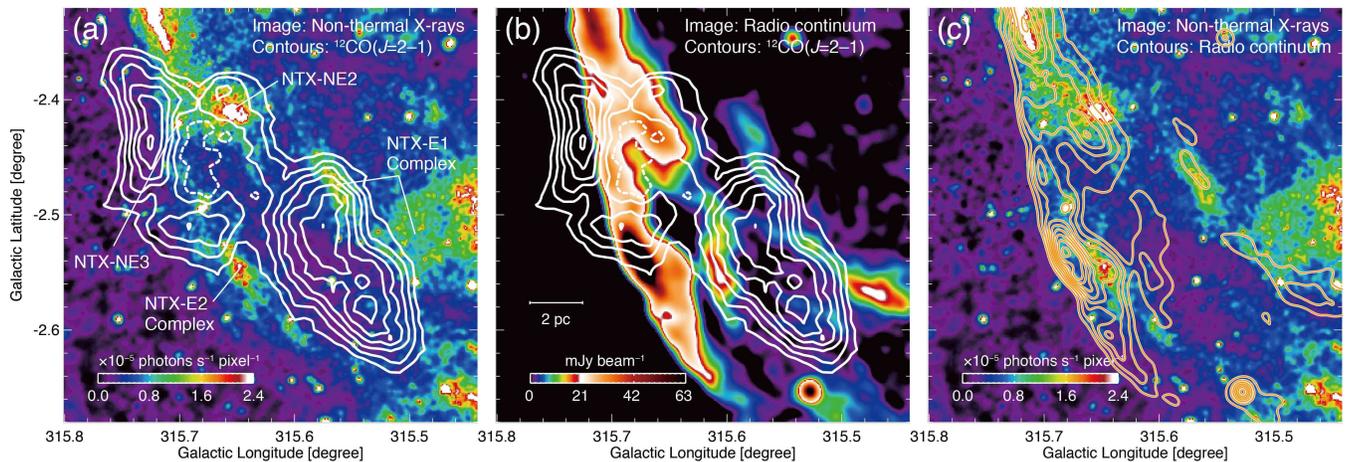}
\caption{(a) Non-thermal X-ray image of RCW 86 in the energy range from 2.0 to 4.5 keV superposed with $^{12}$CO($J$ = 2--1) intensity contours (white) toward the east of the SNR. The color scale is linear, and is given in units of $10^5$ photons s$^{-1}$ pixel$^{-1}$. The CO velocity range is the same as Figure \ref{f06} East. The lowest contour level and the contour interval of CO are 2.1 K ($\sim 15 \sigma$) and 1.0 K ($\sim 10 \sigma$), respectively. (b) Radio continuum image superposed on $^{12}$CO($J$ = 2--1) intensity contours (white). The linear color scale is given in mJy beam$^{-1}$. The contour levels are the same as in the left panel. (c) Same non-thermal X-ray image as in Figure \ref{f14}a, superposed with radio continuum contours (orange). The lowest contour level and the contour interval of the radio continuum are 7 mJy beam$^{-1}$ and 10 mJy beam$^{-1}$, respectively.}
\label{f14}
\end{center}
\end{figure*}%

Within the gas-poor/diffuse region at the NE, the
acceleration by the fast shock is highly efficient. Figure \ref{f09} also shows that the ISM density 
at the NE is lower than at the SW and the N, as shown by the lower H{\sc i} intensity.  H$\alpha$ and X-ray observations indicate that the maximum shock velocity at the NE is $\sim3,000$ km s$^{-1}$ (with an 
average of $\sim1,200$ km s$^{-1}$), 3--6 times larger than that at the SW and NW \citep{1990ApJ...358L..13L,2001ApJ...547..995G,2013MNRAS.435..910H}. The difference in velocity by a factor of 3 corresponds to an $\varepsilon_0$ larger by
an order of magnitude. We thus suggest that in the NE, the fast 
shock waves increased $\varepsilon_0$ and the intensity of the synchrotron radiation. It is suggested that the shock velocity at the SW has been slowing down rapidly for the last 200 years \citep{2013MNRAS.435..910H}. This is consistent with the deformation of the shock front toward NTX-NE1 along the curved H{\sc i} cavity-wall (see Northeast in Figure \ref{f09}). In spite of that, the shock velocity remains three times higher than in the SW, suggesting that the 
H{\sc i} gas is physically associated with the SNR. In 1,000 years we would expect that the shock waves in the NE will come into contact completely with the H{\sc i} cavity wall and the X-rays will be enhanced by the shock-cloud interaction.

\begin{figure*}
\figurenum{A.1}
\begin{center}
\includegraphics[width=174mm,clip]{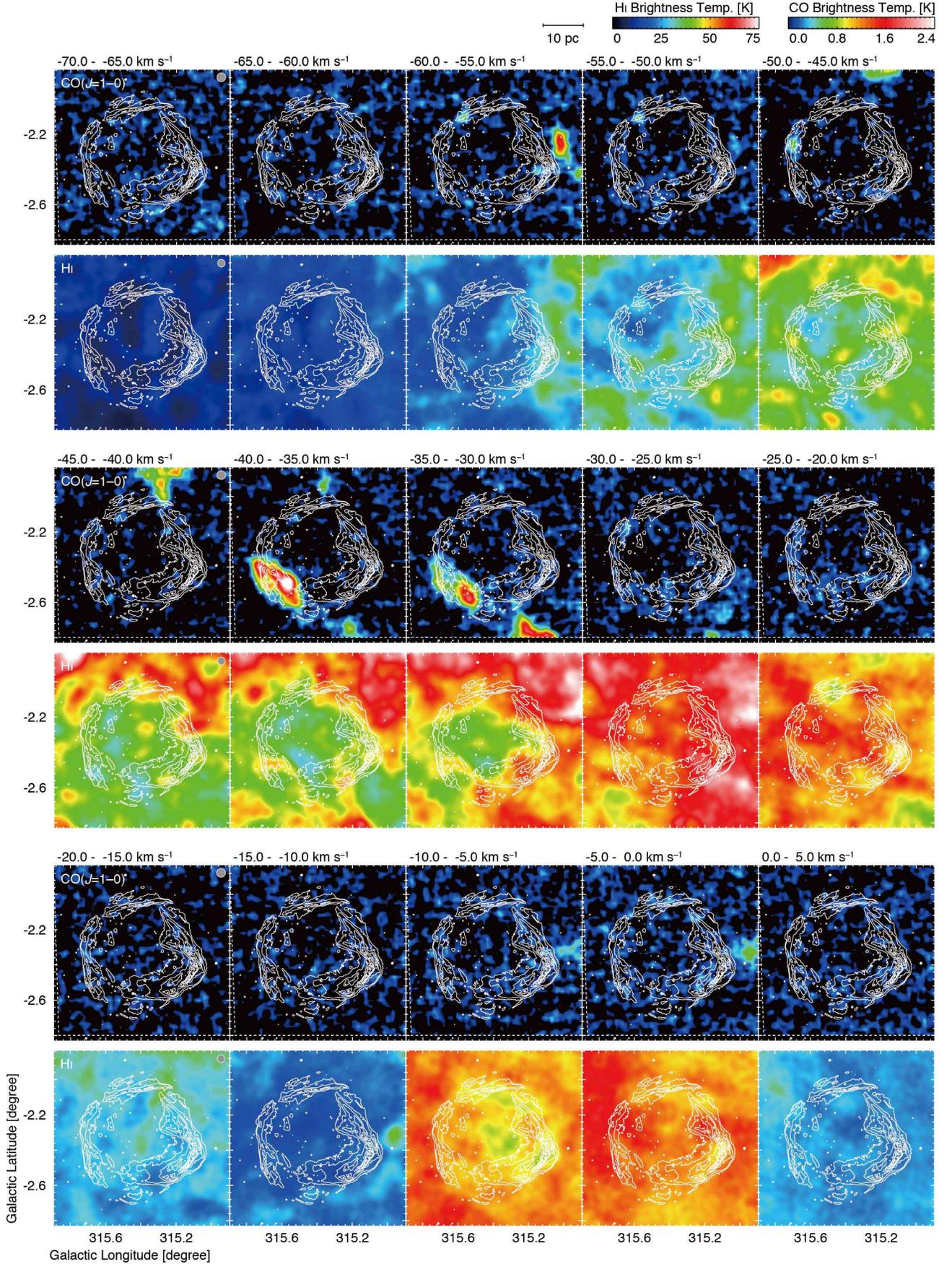}
\caption{Velocity channel distributions of the $^{12}$CO($J$ = 1--0) and H{\sc i} brightness temperatures superposed with the same X-ray intensity contours as in Figure \ref{f03}a. Each panel of CO/H{\sc i} shows intensity distributions averaged every 5 km s$^{-1}$ in a velocity range from $-70$ to +5 km s$^{-1}$. The scale and color bars for H{\sc i} and CO are shown on top of the set of panels.}
\label{fa1}
\end{center}
\end{figure*}%

\subsection{Forward and Reverse Shocks}
In this section we discuss the forward and reverse shocks in RCW 86. Based on $Chandra$ observations toward the SW shell, \cite{2002ApJ...581.1116R} showed that relativistic CR electrons are accelerated by the reverse shock, since the non-thermal X-rays are located at the interior of the heated H{\sc i} traced by thermal X-rays. In addition, 
the different spatial distributions of radio continuum emission and non-thermal X-rays reflect the different energy ranges of the emitting CR electrons. For a magnetic field of 10 $\mu$G, the synchrotron radiation whose peak is $h\nu$ = 4 keV loses the energy at a decay timescale of only 900 years, while the CR electrons emitting at 1 GHz radio continuum can radiate over $10^7$ years. The synchrotron X-rays originate in
the high-energy electrons close to the shock front and the radio emission from 
lower energy electrons downstream \citep{2002ApJ...581.1116R}. \cite{2007PASJ...59S.171U} and \cite{2011PASJ...63S.837Y} studied the Fe ejecta distribution over the SNR and showed that highly ionized Fe ejecta are likely heated up by the reverse shock. 

Our comparative study of multi-wavelength observations of the ISM provides a tool to discriminate the reverse shock and the forward shock in different regions. In Figure \ref{f11} we showed the relative location between the non-thermal X-ray peaks and radio peaks in the radial distribution. The positive values of the separation (Separation $> 0$) correspond to the case in which the non-thermal X-ray peaks are located farther inside than the nearest radio peak, while the negative values of the separation (Separation $< 0$) correspond to the opposite case. Rho et al. (2002) interpreted that the former case corresponds to the reverse shock and the latter, to the forward shock.

The reverse shock is located only in regions with the average gas column density $<$$N_\mathrm{p}(\mathrm{H}_2 + \mathrm{H}{\textsc{i}})$$>$ $> 10^{21}$ cm$^{-2}$, whereas the forward shock is located, except for one point, in regions with the average gas column density $<$$N_\mathrm{p}(\mathrm{H}_2 + \mathrm{H}{\textsc{i}})$$>$ $< 10^{21}$ cm$^{-2}$. This supports the idea that the reverse shock recoiled after collision with the ISM. It is possible that the exceptional point be explained if the shock slipped through the clumpy gas-rich region. Conversely, in the gas-poor region the reverse shock has not been detected. 
Particularly, toward the dense clumpy CO $-37$ E the forward/reverse shock has a complicated distribution. Figure \ref{f14} shows the X-ray
non-thermal and radio continuum distribution of CO $-37$ E. The upper right corner of the images points in the direction of the SNR center. In addition to the NTX-NE2--3 complexes, we find various non-thermal X-ray filaments similar to NTX-E1--2 as a representative case and the filaments show complimentary distributions to the clumpy CO clouds. We see a trend consisting in the non-thermal X-rays being located more in the inner part than the radio continuum. In the typical region NTX-NE2--3, the separation between the non-thermal and radio continuum is 48--64 arcsec{,} which is equivalent to 0.6--0.8 pc at 2.5 kpc. The shock speed of this area is estimated to be $1,653 \pm 228$ km s$^{-1}$ by \cite{2013MNRAS.435..910H}. If we assume that the reverse shock and the forward shock are moving at the same velocity, the shock wave collided with the molecular cloud 400 yrs ago. This agrees well with the shock age $< 380$ yr of Fe ejecta \citep[e.g.,][]{2008PASJ...60S.123Y}.

In addition, the reverse shock may hold the key to understand an efficient acceleration mechanism of CR electrons with $\sim$1 TeV or higher. According to the numerical simulations, turbulence in downstream regions can create strong magnetic fields, up to mG or $\sim$ 50 $\mu$G on average \citep[e.g.,][]{2012ApJ...744...71I}. In this unusual situation, some additional acceleration mechanisms will become important, including acceleration with magnetic reconnection \citep[e.g.,][]{2012PhRvL.108m5003H}, reverse shock acceleration \citep[e.g.,][]{2001SSRv...99..305E}, non-linear effect of DSA \citep[e.g.,][]{2001RPPh...64..429M}, second-order Fermi acceleration \citep{1949PhRv...75.1169F}, etcetera. Detailed X-ray spectroscopy and comparative studies with the interstellar gas will reveal efficient acceleration mechanisms of CR electrons.

To summarize, a thorough investigation of the ISM is extremely important to study the progenitor system, the origin of thermal X-rays, the acceleration mechanism of CR electrons, and the shock dynamics of SNRs. Observations of the ISM at high angular resolution better than 45 arcsec will allow us to make a comparison of small-scale structures of the ISM with observations at the other wavelengths, and will enable us to pursue more detailed physical process. A spectral analysis of the X-ray data is indispensable to derive the distributions of the photon index and the roll-off energy and provide a firm basis to elucidate the relationship between the CR acceleration and the ISM. The synchrotron radiation above 10 keV from the electrons accelerated by the reverse shock will be obtained in the hard X-ray imaging with $Nustar$. $Chandra$ high-resolution measurements of the proper motion will reveal the kinematics of the X-ray filaments in detail.

\section{CONCLUSIONS}
We summarize the present work as follows.

\begin{enumerate}
\item We have revealed atomic and molecular gas associated with the young TeV $\gamma$-ray SNR RCW 86 by using NANTEN2 CO and ATCA $\&$ Parkes H{\sc i} datasets. The H{\sc i} gas is distributed surrounding the X-ray shell and shows a cavity-like distribution with an expanding velocity of $\sim7$ km s$^{-1}$, while the CO clouds are located only in the east, south, and northwest, showing the high-intensity ratio of CO $J$ = 2--1/1--0 ratio $> 0.8$ enhanced by the shock heating and/or compression in the surface of the clouds.
\item Thermal X-ray filaments show a good spatial correspondence with the H{\sc i} wall and small-scale structures of CO clouds. We also found a correlation between the total proton column density and the thermal X-ray intensity. This indicates that the atomic/molecular gas of density 10--100 cm$^{-3}$ is associated with the SNR shockwaves.
\item Non-thermal X-rays are bright both in the gas-rich and -poor regions. We interpret that the shock-cloud interaction between the cold H{\sc i} clumps and the high shock velocity could enhance the non-thermal X-rays, which is a situation similar to that discussed by \cite{2015ApJ...799..175S} in the SNR RX J1713.7$-$3946. In addition, the reverse shock is detected only in the gas-rich region with a total proton column density of $\sim10^{21}$ cm$^{-2}$ or higher.
\item Our study confirms that the progenitor of RCW 86 was a system consisting of a white dwarf and a low-mass star with low-velocity accretion winds, as suggested by \cite{2011ApJ...741...96W}.
\end{enumerate}

\section*{ACKNOWLEDGEMENTS}
We are grateful to Aya Bamba, Takaaki Tanaka, Hiroyuki Uchida, and Hiroya Yamaguchi for thoughtful comments and their contribution on the X-ray properties. We acknowledge Anne Green for her valuable support during the H{\sc i} observations and reduction, and Gloria M. Dubner, PI of the ATCA project C1011 carried out to obtain the reported H{\sc i} data, who provided them to Yasuo Fukui. We also acknowledge to Shinya Tabata, Momo Hattori, Shigeki Shimizu, Sho Soga, Daichi Nakashima, Shingo Otani, Yutaka Kuroda, Masashi Wada, Ryo Kaji, Keisuke Hasegawa, and Rey Enokiya for contributions on the observations of $^{12}$CO($J$ = 1--0) data. This work was financially supported by Grants-in-Aid for Scientific Research (KAKENHI) of the Japanese society for the Promotion of Science (JSPS, grant Nos. 22740119, 12J10082, 24224005, 15H05694, and 16K17664). This work also was supported by ``Building of Consortia for the Development of Human Resources in Science and Technology'' of Ministry of Education, Culture, Sports, Science and Technology (MEXT, grant No. 01-M1-0305). This research was based on observations obtained with $XMM$-$Newton$, an ESA science mission with instruments and contributions directly funded by ESA Member States and NASA. We also utilize data from MOST and SHASSA. The Molonglo Observatory Synthesis Telescope (MOST) is operated by The University of Sydney with support from the Australian Research Council and the Science Foundation for Physics within The University of Sydney. The Southern H-Alpha Sky Survey Atlas (SHASSA) is supported by the National Science Foundation. EMR is member of the Carrera del Investigador Cient\'\i fico of CONICET, Argentina, and is partially supported by CONICET grant PIP 112-201207-00226.

\section*{APPENDIX: VELOCITY CHANNEL MAPS OF CO AND H{\sc i}}
Figure \ref{fa1} shows the velocity channel distributions of the $^{12}$CO($J$ = 1--0) and H{\sc i} brightness temperature every 5 km s$^{-1}$ from $-70$ km s$^{-1}$ to +5 km s$^{-1}$ superposed with the X-ray intensity contours. First, we investigated the spatial correlation and anti-correlation between the X-ray and interstellar gas (CO and H{\sc i}) distributions. We found that the X-ray shell is complementary to the CO/H{\sc i} structure at a radial velocity of $\sim -35$ km s$^{-1}$. In particular, the H{\sc i} cavity and its expanding motion showed evidence for the association with the SNR shocks (see also Figure \ref{f05} and Section 3.3). \cite{2016ApJ...819...98A} presented the image of the H{\sc i} cavity-like structure, but the authors did not mention the existence of an expanding shell motion. Apparently, the H{\sc i} cloud at a radial velocity from $-10$ to 0 km s$^{-1}$ is also well correlated with the X-ray shell; however, as it is a local cloud component, we ignored it from the standpoint of interaction with the SNR.

\end{document}